\documentclass[aip,jcp,amsmath,amssymb,reprint]{revtex4-1}

\usepackage{graphicx}
\usepackage{graphicx,color}
\usepackage{dcolumn}
\usepackage{bm}
\usepackage{ulem}

\begin{document}

\preprint{AIP/123-QED}

\title{Entropy of self-avoiding branching polymers: Mean-field theory and Monte Carlo simulations}

\author{Davide Marcato}
\email{dmarcato@sissa.it}
\affiliation{Scuola Internazionale Superiore di Studi Avanzati (SISSA), Via Bonomea 265, 34136 Trieste, Italy}
\author{Achille Giacometti}
\email{achille.giacometti@unive.it}
\affiliation{Dipartimento di Scienze Molecolari e Nanosistemi, Universit\`a Ca' Foscari Venezia, 30123 Venezia, Italy}
\affiliation{European Centre for Living Technology (ECLT) Ca' Bottacin, 3911 Dorsoduro Calle Crosera, 30123 Venezia, Italy}
\author{Amos Maritan}
\email{amos.maritan@pd.infn.it}
\affiliation{Laboratory of Interdisciplinary Physics, Department of Physics and Astronomy ``G. Galilei'', University of Padova, Padova, Italy and INFN, Sezione di Padova, via Marzolo 8, 35131 Padova, Italy}
\author{Angelo Rosa}
\email{anrosa@sissa.it}
\affiliation{Scuola Internazionale Superiore di Studi Avanzati (SISSA), Via Bonomea 265, 34136 Trieste, Italy}


\date{\today}

\begin{abstract}
We study the statistics of branching polymers with excluded-volume interactions, by modeling them as single self-avoiding trees on a generic regular periodic lattice with coordination number $q$. Each lattice site can be occupied at most by one tree node, and the fraction of occupied sites can vary from dilute to dense conditions.
By adopting the statistics of rooted-directed trees as a proxy for that of undirected trees without internal loops and by an exact mapping of the model into a field theory, we compute the entropy and the mean number of branch-nodes within a mean-field approximation and in the thermodynamic limit.
In particular, we find that the mean number of branch-nodes is independent of both the lattice details and the lattice occupation, depending only on the associated chemical potential.
Monte Carlo simulations in $d=2,3,4$ provide evidence of the remarkable accuracy of the mean-field theory, more accurate for higher dimensions.
\end{abstract}

\maketitle

{\it Introduction} --
The physical behavior of many polymeric systems can be ascribed to the fact that their smallest constituents assume highly branched conformations, with no internal loops~\cite{FisherPRL1978,KurtzeFisherPRB1979,IsaacsonLubensky,DuarteRuskin1981,ParisiSourlasPRL1981,BovierFroelichGlaus1984,Burchard1999,RubinsteinColby,Rosa2016a,Rosa2016b,Rosa2016c,Ghobadpour2021,VanDerHoek2024}.
In soft matter physics, for instance, tree-like structures emerge as the consequence of topological constraints between non-concatenated and unknotted circular polymers in dense phases~\cite{KhokhlovNechaev85,RubinsteinPRL1986,RubinsteinPRL1994,GrosbergSoftMatter2014,RosaEveraersPRL2014,MichielettoSoftMatter2016,RosaEveraers2019,SchramSM2019,SmrekRosa2019}.
In biology, branched structures describe the behavior of fundamental biomolecules like
supercoiled bacterial DNA~\cite{MarkoSiggia1994,MarkoSiggiaSuperCoiledDNA1995,odijk_osmotic_1998,Cunha2001,Junier2023},
chromosomal DNA during interphase~\cite{grosbergEPL1993,RosaPLOS2008,Vettorel2009,halverson2014melt}
and viral RNA~\cite{GelbartPNAS2008,Fang2011,Vaupotic2023}.
A distinctive feature of all these examples is the fact that branching is {\it annealed}, namely the positions of branch-nodes on the molecular backbone fluctuate~\cite{GutinGrosberg93,CuiChenPRE1996,Rosa2016a} and are in thermal equilibrium with the other conformational degrees of freedom of the system.
The entropic contribution arising from branching was studied mainly in the context of ideal molecules, namely disregarding any form of monomer-monomer interaction~\cite{ZimmStockmayer49,DeGennes1968,DaoudJoanny1981,GrosbergNechaev2015,EveraersGrosbergRubinsteinRosa,VanDerHoek2025}.
More realistic models, with the introduction of explicit excluded-volume interactions, become theoretically challenging and, in fact, only very few exact results are available~\cite{LubenskyIsaacson1979,Lubensky1981,ParisiSourlasPRL1981}.

Approximate theories were proposed in the past to study interacting branching polymers~\cite{LubenskyIsaacson1979,Lubensky1981,Nemirovsky1992,Gujrati1995,Banchio1995,deLosRios2000,Safran2002,Wagner2015}.
In most cases~\cite{LubenskyIsaacson1979,Lubensky1981,Nemirovsky1992,Safran2002,Wagner2015} they consist in mean-field (hereafter, MF) approximations of lattice field theories constructed upon the ``$n \to 0$''-limit of the spin $O(n)$-model of a magnet, following the original idea for linear chains proposed first by Daoud {\it et al.}~\cite{Daoud1975} and formalized later by de Gennes~\cite{DeGennesBook}.
Yet, these classical works study systems where the presence of loops cannot be truly avoided, and the number and size of the polymer molecules cannot be controlled.
Another possible approximation consists in embedding the polymer system into a Bethe lattice~\cite{Gujrati1995,Banchio1995}.
In this framework, the absence of loops in the polymer structure is ensured by the loopless nature of the underlying lattice itself.
However, the constraint that a polymer in real space (or in a conventional periodic lattice) must avoid loops introduces a nontrivial correlation between nodes, which is neglected in the Bethe lattice representation.
An alternative attempt to construct a theory for single-chain systems avoiding loops on a standard periodic lattice was proposed in~\cite{deLosRios2000}: here, however, the corresponding MF free energy requires numerical minimization and the amount of branch-nodes in the system cannot be controlled, keeping into the same ensemble configurations of linear chains and highly-branched structures.

Motivated by these considerations, in this Communication we introduce a model for branching polymers with {\it explicit} excluded-volume interactions which solves all these issues, namely:
(i) loops are rigorously excluded, thus providing a general framework for naturally occurring systems like those mentioned at the beginning;
(ii) the approach is genuinely single-chain;
(iii) control is maintained on both the amount of branch-nodes (and, so, explore the whole range from the poorly- to the highly-branched regime) and monomer concentration.
We then proceed via an {\it exact} mapping of the partition function on a field theory which can be solved within a mean-field approximation.
This solution includes many other known results as limit cases and allows us to derive a transparent analytical formula for the mean number of branch-nodes of the trees which is in semi-quantitative agreement with known predictions from exact-enumeration techniques and new Monte Carlo computer simulations of self-avoiding trees on the hypercubic lattice in $d=2,3,4$ spatial dimensions.

\begin{figure}
\includegraphics[width=0.40\textwidth]{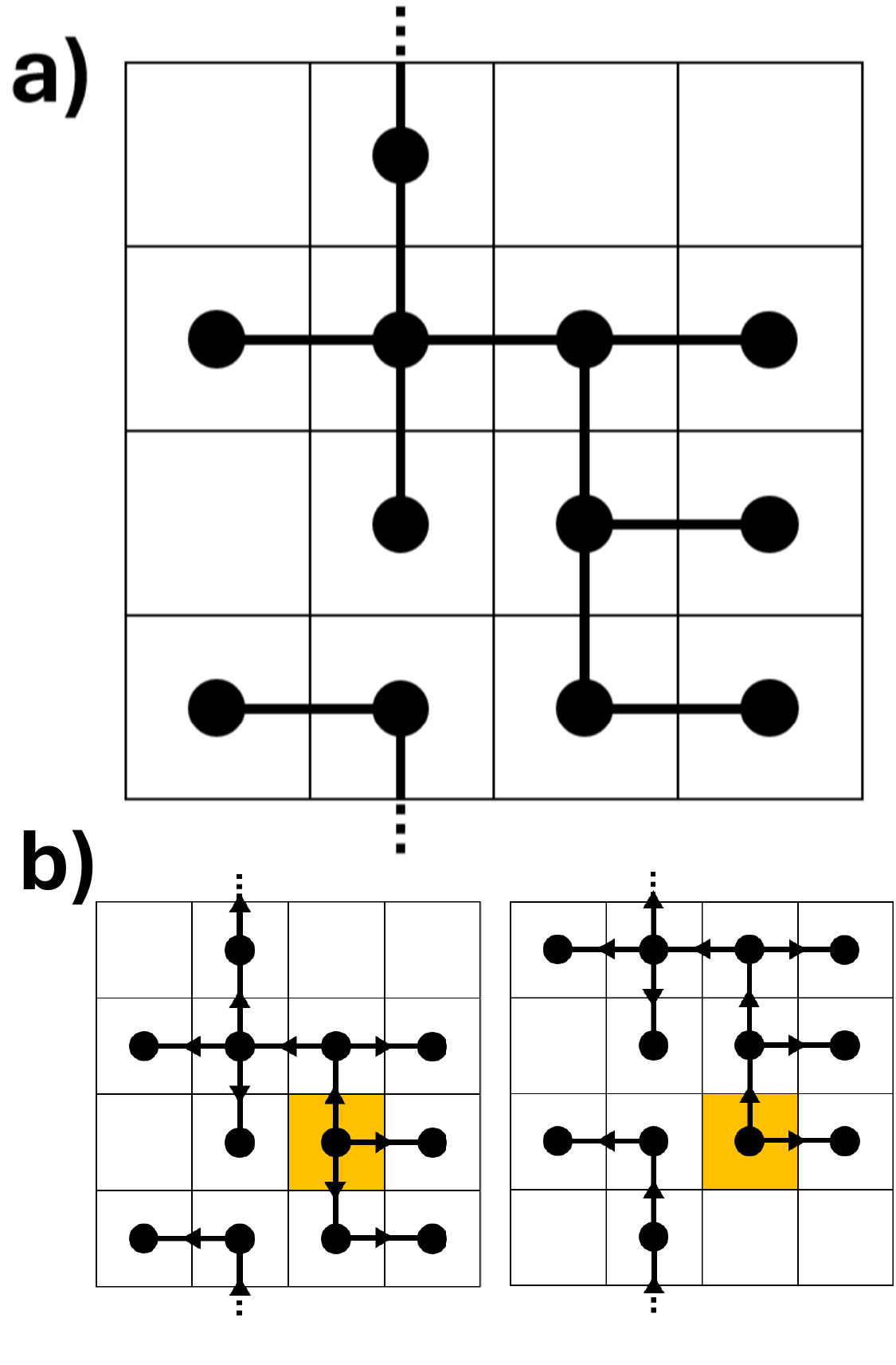}
\caption{
(a)
Illustration of a SAT on the square lattice ($q=4$, $V=4\times 4=16$ total sites) with $N=12$, $N_1=6$, $N_2=3$, $N_3=2$ and $N_4=1$.
Notice that $N_1$ and $N_2$ verify Eqs.~\eqref{eq:N1} and~\eqref{eq:N2}.
The dashed bond illustrates the role of periodic boundary conditions (p.b.c.).
(b, left)
The same polymer conformation of panel (a) as a rooted-directed SAT, with the root site marked in orange.
(b, right)
A different rooted-directed SAT with the root on the same lattice site and the same connectivity structure (through p.b.c.) of the conformation on the left.
}
\label{fig:ExampleConformation}
\end{figure}
%

{\it The model} --
We consider a regular lattice with coordination number $q$, total number of sites $V$, and where periodic boundary conditions are assumed.
A single randomly branching polymer with excluded-volume interactions is modeled as a self-avoiding tree (SAT) made of $N \leq V$ monomers (nodes), or $N-1$ bonds, on such lattice.
A site of the lattice can be occupied by at most one tree node and the bond between two nodes (equal to one lattice step) corresponds to our length unit (see Fig.~\ref{fig:ExampleConformation}(a) for a simple SAT conformation on the square lattice). 
In the absence of loops, denoting with $f_i$ the number of bonds emanating from node $i$ (the so called {\it functionality}) and with $N_f$ the total number of nodes of functionality $f$, we have that $N = \sum_{f=1}^q N_f$ while the $f_i$'s sum to twice the total number of bonds, {\it i.e.} $\sum_{i=1}^N f_i = \sum_{f=1}^q f \, N_f = 2(N-1)$.
Then, from these relations one obtains:
\begin{eqnarray}
N_1 & = & \sum_{f = 3}^q (f-2) N_f + 2 \, , \label{eq:N1} \\
N_2 & = & N - \sum_{f = 3}^q (f-1) N_f -2 \, , \label{eq:N2}
\end{eqnarray}
where the upper limit $f=q$ in the sums follows from the self-avoidance condition.\footnote{Of course, for ideal trees with no excluded volume nothing prevents $f$ to be $>q$. Notice also that the obvious relation $N=\sum_{f=1}^q N_f$ follows immediately from Eqs.~\eqref{eq:N1} and~\eqref{eq:N2}.}

{\it SAT entropy, MF approximation} --
Our task consists in computing $\Omega_N( \left\{ N_f \right\})$, the total number of SAT's of $N$ nodes and $\left\{ N_f \right\} \equiv (N_3, ..., N_q)$ branch-nodes.
To this purpose, we consider the related problem of the so called {\it rooted-directed} SAT's, namely trees where every bond has an assigned orientation such that there exists a unique directed path from a ``privileged'' node (called {\it root}),
whose spatial position is taken at the (arbitrarily chosen) {\it fixed} lattice site position ${\mathbf x}_0$ (see panel (b) in Fig.~\ref{fig:ExampleConformation}).
Then, the total number of rooted-directed SAT's, $\Omega_N^{{\mathbf x}_0}(\left\{ N_f \right\})$, is related to $\Omega_N(\left\{ N_f \right\})$ as shown by the following argument.
As a direct consequence of (i) assuming periodic boundary conditions and (ii) fixing the root position ${\mathbf x}_0$, each distinct tree conformation has a degeneracy due to translations equal to $V$ in the case of undirected configurations, and equal to $N$ in the case of rooted-directed configurations.
In fact, for a rooted-directed SAT configuration there must always be a monomer (\textit{i.e.} the root) in position $\bold{x}_0$, thus limiting the possible translations.
The degeneracy due to rotations is instead the same for both undirected and rooted-directed configurations (details, in particular Fig.~\ref{fig:TreeConformationsUNdirectedDirected}, in Sec.~\ref{sec:UndirectedDirectedTrees} in Supplementary Material (SM)).
Hence the ratio $\Omega_N(\left\{ N_f \right\}) / \Omega_N^{{\mathbf x}_0}(\left\{ N_f \right\}) = V/N$. 
Then, by introducing the Boltzmann constant $k_B$ we get the relation:
\begin{equation}\label{eq:DirectedUndirectedRelation}
s = \frac{k_B}V \log \left( \frac{V}N \right) + s^{{\mathbf x}_0}
\end{equation}
between the entropies per lattice site, $s \equiv k_B \log(\Omega_N) / V$ and $s^{{\mathbf x}_0} \equiv k_B \log(\Omega_N^{{\mathbf x}_0}) / V$, for undirected and rooted-directed SAT's respectively.
In the thermodynamic limit $V \gg 1$ we can neglect terms of order $1/V$, and Eq.~\eqref{eq:DirectedUndirectedRelation} implies $s \simeq s^{{\mathbf x}_0}$. 
This observation is now crucial because, contrary to $\Omega_N$, $\Omega_N^{{\mathbf x}_0}$ can be mapped {\it exactly} on a field theory,\footnote{Essentially, the reason why we get ``easily'' $\Omega_N^{{\mathbf x}_0}$ but not $\Omega_N$ is because by assigning a direction to each bond, looped conformations have two bonds pointing to the same site: in this way (see Sec.~\ref{sec:FieldTheoryCompleteDerivation} in SM) their statistical weight is $=0$ and are so naturally removed from the count. At the same time, by our rules it is not too difficult to demonstrate (Sec.~\ref{sec:CommentOnZandiWork} in SM) that an alternative field theory like the one proposed by Wagner {\it et al.}~\cite{Wagner2015} does, in fact, count loops.} for which we can adopt a scheme similar to that recently introduced  by us~\cite{Marcato2023,Marcato2024} to model interacting linear polymer solutions and largely based on the well-known de Gennes' mapping of the ``$n \to 0$''-limit of the spin $n$-vector model~\cite{DeGennesBook} (the complete derivation of $\Omega_N^{{\mathbf x}_0}$ is given in Sec.~\ref{sec:FieldTheoryCompleteDerivation} in SM).
The main idea consists in writing the {\it grand canonical} partition function of the system $\Xi$ in the following field-theoretic form,
\begin{equation}
\Xi = \lim_{n \to 0} \int {\mathcal D}{\bm \varphi} \, {\mathcal D}\bar{\bm \varphi} \, e^{-{\mathcal A}({\bm \varphi}, \bar{\bm \varphi})} \, ,
\end{equation}
where $\bm \varphi$ and $\bar{\bm \varphi}$ are $n$-component field vectors and ${\mathcal A}({\bm \varphi}, \bar{\bm \varphi})$ is an effective action (Eq.~\eqref{eq:FieldTheoreticZ} in SM, whose validity can be further assessed by applying a version of the famous Wick's theorem~\cite{ZinnJustin2021} to it (see Sec.~\ref{sec:UseWickTheorem} in SM)).
While the final expression remains mathematically challenging, it is particularly amenable to an expansion around the saddle-point~\cite{Orland1985,Orland1996,Higuchi1998,Marcato2023,Marcato2024}, which is obtained by minimizing the effective action.
The first term of this expansion corresponds to our MF estimate.
Finally, a standard Legendre transform allows us to switch back to the canonical ensemble and get finally $\Omega_N^{{\mathbf x}_0}$.
By neglecting all terms up to ${\mathcal O}(V^{-1})$, the MF expression for $s = s_{\rho}(\left\{ \phi_f \right\})$ (Eq.~\eqref{eq:DirectedUndirectedRelation}) for finite node density $\rho \equiv N/V$ and as a function of branch-node fractions $\left\{ \phi_f \right\} \equiv (\phi_3, ..., \phi_q)$ with $\phi_f \equiv N_f/N$ is given by 
\begin{eqnarray}\label{eq:MFFreeEnergy}
\frac{s_{\rho}(\left\{ \phi_f \right\})}{k_B}
& \simeq & \rho \log \bigg(\frac{q}{e}\bigg) - (1-\rho)\log(1-\rho) \nonumber\\
& & - \rho \sum_{f=1}^q \phi_f \log( \phi_f(f-1)! ) \, ,
\end{eqnarray}
with $e = 2.718...$ being the Euler number, $(...)!$ the factorial symbol and with the constraints $\phi_1 =  \sum_{f = 3}^q (f-2) \phi_f$ and $\phi_2 = 1 - \sum_{f = 3}^q (f-1) \phi_f$ following, respectively, from Eqs.~\eqref{eq:N1} and~\eqref{eq:N2}.

{\it Validation, limiting behaviors} --
The central formula Eq.~\eqref{eq:MFFreeEnergy} recapitulates a number of important results as limiting cases.
First, in the particular case of linear chains where $N_1=2$, $N_2=N-2$ and $N_{f\geq 3}=0$ ({\it i.e.} $\phi_1 \simeq0$, $\phi_2\simeq 1$ and $\phi_{f\geq 3}=0$) Eq.~\eqref{eq:MFFreeEnergy} reduces to $s = \rho \log \left(\frac{q}{e}\right) - (1-\rho)\log(1-\rho)$, corresponding to the accurate MF approximation of Orland and coworkers for Hamiltonian and self-avoiding walks~\cite{Orland1985,Orland1996}.
Second, in the dilute regime with $\rho \ll 1$, to linear order in $\rho$ Eq.~\eqref{eq:MFFreeEnergy} reads
\begin{equation}\label{eq:EntropyDiluted}
\frac{s_{\rho}(\left\{ \phi_f \right\})}{k_B} \simeq \rho \log \bigg( q \prod_{f=1}^q \frac1{(\phi_f (f-1)!)^{\phi_f}} \bigg) \, ,
\end{equation}
{\it i.e.} the entropy per node $s_{\rho}/\rho$ is constant.
Interestingly, the argument of the logarithm plays now the role of an effective coordination number~\cite{Lipson1983} which depends on the set of branching probabilities $\{\phi_f\}$.
In particular, this effective coordination number can take {\it smaller} values with respect to the one for linear chains (which, within our MF approximation, is $= q$, {\it i.e.} the same as for random walks on the lattice): for instance, by setting $\phi_{3\leq f < q}=0$ and $\phi_q \neq 0$, for $q \geq 6$, there are values of $\phi_q$ giving an effective coordination number $<q$.
Third, of even greater interest is the opposite limit of {\it spanning} trees, $\rho=1$, namely at full lattice occupation.
In particular, through the Legendre transform of Eq.~\eqref{eq:MFFreeEnergy} (see Sec.~\ref{sec:EnumerationSpanningTrees} in SM) it is possible to enumerate the {\it total} number of spanning trees on the lattice with {\it unrestricted} $N_f$.
Our MF estimate reproduces the known~\cite{Wu1977,Shrock2000} exact entropies of spanning trees for different lattices with good accuracy (deviations are no larger than $\simeq 20\%$, see Table~\ref{tab:ComparisonExactMeanField} in SM). 
While our approximation is less precise than the Bethe approximation of De Los Rios {\it et al.}~\cite{deLosRios2000} (see Table~\ref{tab:ComparisonExactMeanField} in SM), our method is conceptually superior as it allows us to seize control of the mean number of branch-nodes, a feature absent in~\cite{deLosRios2000}. 

Finally, it is interesting to compare Eq.~\eqref{eq:MFFreeEnergy} with $\rho=1$ to the asymptotic behavior of the exact graph-theoretical expression presented in Ref.~\cite{VanDerHoek2025},
\begin{equation}\label{eq:PieterS}
\frac{s_{\rm ideal}(\left\{ \phi_f \right\})}{k_B} \simeq \log \bigg(\frac{N}e\bigg) -\sum_{f=1}^{N-1} \phi_f \log[\phi_f(f-1)!] \, ,
\end{equation}
valid for trees with no excluded-volume.
The leading term in Eq.~\eqref{eq:PieterS} implies that $s_{\rm ideal} \gg s_{\rho=1}$ for $N\gg1$, as one would intuitively expect.
At the same time, and quite remarkably, the two sums in Eq.~\eqref{eq:EntropyDiluted} and Eq.~\eqref{eq:PieterS} are fundamentally identical were not for their different upper limit reflecting, respectively, the excluded-volume and the ideal condition. 
We will comment more on this issue in the next point, in the context of branch-node statistics.

\begin{figure}
\includegraphics[width=0.45\textwidth]{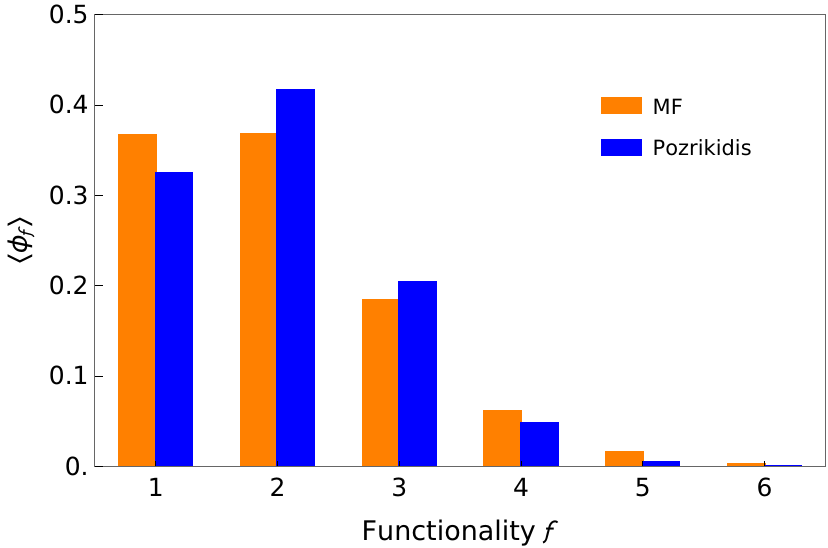}
\caption{
Distribution of equilibrium densities $\langle \phi_f\rangle$ for nodes of functionality $f$ in spanning trees on the triangular lattice ($q=6$).
Orange columns correspond to our MF result for an infinite lattice, whereas blue columns refer to the results obtained through the numerical procedure of Pozrikidis~\cite{Pozrikidis2016} for a periodic lattice with a total number of $256$ sites.
}
\label{fig:FunctionalityDistribution}
\end{figure}
\begin{figure*}
$$
\begin{array}{ccc}
\includegraphics[width=0.32\textwidth]{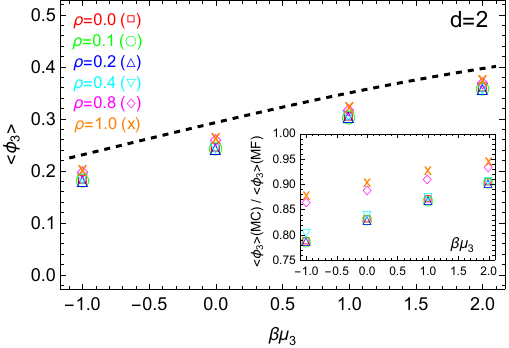} &
\includegraphics[width=0.32\textwidth]{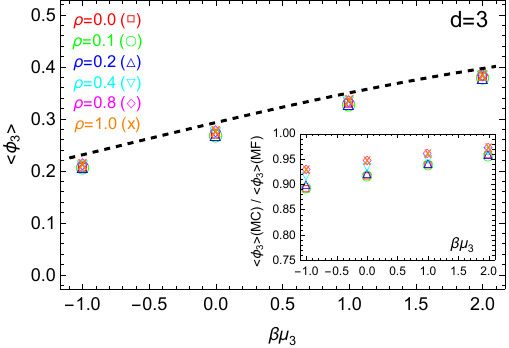} &
\includegraphics[width=0.32\textwidth]{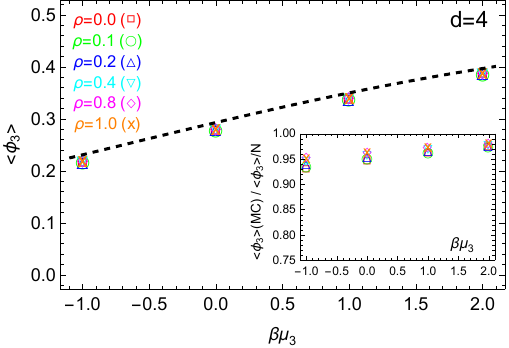}
\end{array}
$$
\caption{
Asymptotic mean fraction of branch-nodes, $\langle \phi_3\rangle$, as a function of the branch chemical potential $\beta\mu_3$, comparison between MF prediction Eq.~\eqref{eq:BranchingProbability} (dashed line) and MC numerical simulations (symbols, corresponding to the values for the largest $N$'s reported in Fig.~\ref{fig:2Ddata} in SM) for different node densities $\rho$ (see colorcode/symbol in the legend) and in spatial dimensions $d=2,3,4$.
(Insets)
Corresponding ratios between MC results and the MF prediction.
}
\label{fig:<n3>/N}
\end{figure*}
%

{\it Branch-nodes, comparison to numerical results} --
In terms of Eq.~\eqref{eq:MFFreeEnergy}, the chemical potential for the $f$-functional branch-nodes, $\mu_f$, at temperature $T$ is expressed by
\begin{equation}\label{eq:BranchMuf}
\beta\mu_f = -\frac1{k_B \rho} \frac{\partial s_{\rho}(\{\phi_f\})}{\partial \phi_f} \, ,
\end{equation}
where $\beta = 1/(k_BT)$.
Inverting Eq.~\eqref{eq:BranchMuf} for given $\{\mu_f\}$, the unknown mean values $\{\langle \phi_f\rangle\}$ obey a system of $q-2$ equations which, in general, needs to be solved numerically.
However, a general feature easily deduced from Eqs.~\eqref{eq:MFFreeEnergy} and~\eqref{eq:BranchMuf} is that the values of $\{\langle \phi_f\rangle\}$ do not depend on $\rho$.
We study first the case of spanning trees ({\it i.e.} $\rho=1$) with unrestricted functionality on the triangular lattice ({\it i.e.} $q = 6$ and $\beta\mu_f = 0$ for $3\leq f \leq 6$) for which accurate numerical predictions are available~\cite{Pozrikidis2016}.
As shown in Fig.~\ref{fig:FunctionalityDistribution}, our MF theory agrees very well with the numerical results.
Then we consider the situation where $\beta\mu_f \neq 0$ and assume that branch-nodes are at most $3$-functional regardless of lattice coordination $q$. 
With that, Eq.~\eqref{eq:BranchMuf} reduces to the following simple expression for $\langle\phi_3\rangle$ as a function of the chemical potential $\mu_3$:
\begin{equation}\label{eq:BranchingProbability}
\langle \phi_3\rangle = \frac1{2 + e^{-(\beta\mu_3-\log(2))/2}} \, .
\end{equation}
Eq.~\eqref{eq:BranchingProbability} correctly reproduces the expected limiting behaviors of linear chains ($\beta\mu_3\to -\infty$) and maximally branched polymers ($\beta\mu_3\to +\infty$), respectively $\langle \phi_3\rangle \to 0$ and $\langle \phi_3\rangle \to 1/2$.
Noticeably, Eq.~\eqref{eq:BranchingProbability}, which is independent of $q$, is identical to the corresponding (and exact) quantity for ideal trees (see Ref.~\cite{GhobadpourPhDThesis} and Eq.~(13) in~\cite{VanDerHoek2025}) which one can derive easily from Eq.~\eqref{eq:PieterS}.
As Eq.~\eqref{eq:BranchingProbability} is the result of a MF-like approximation for SAT's, it is legitimate to ask how good this approximation is.
To address this issue, we performed Monte Carlo (MC) computer simulations of single SAT conformations of $N$ nodes and different node density $0 \leq \rho \leq 1$ ({\it i.e.} from dilute to spanning-tree conditions), on the hypercubic lattice in $d=2,3,4$ spatial dimensions, or $q=2d=4,6,8$.
For comparison, and to verify the exactness of Eq.~\eqref{eq:BranchingProbability} in this situation, we simulated also ideal trees in $d=3$.
The goal is to sample trees according to the Boltzmann distribution,
\begin{equation}\label{eq:GranCanonicalDistribution}
P_N(N_3) = \frac{e^{\beta\mu_3 N_3}}{\sum_{N_3'} \, e^{\beta\mu_3 N_3'} \, \Omega_N(N_3')} \, ,
\end{equation}
for tree conformations with $N_3$ branch-nodes, weighed through the chemical potential $\beta\mu_3$. 
For node density $\rho<1$, we use the simple and efficient Amoeba algorithm~\cite{SeitzKlein1981,Rosa2016a,Rosa2016b,VanDerHoek2024}: one of the nodes with $f=1$ is picked at random, removed from the tree and reattached to either one of the ``$f=1$''- or ``$f=2$''-nodes (to preserve the global condition $f\leq 3$), provided that the self-avoiding condition is satisfied.
At full lattice occupation ($\rho=1$), where Amoeba is clearly unsuitable, classical algorithms for spanning trees~\cite{Broder1989,Aldous1990} also fail in the present context because they do not allow to weigh conformations according to Eq.~\eqref{eq:GranCanonicalDistribution}.
To solve this issue, we propose the following new method (for more details, see Sec.~\ref{sec:MCalgos} in SM): starting from an initial spanning tree, a ``$f=1$''-node is randomly picked and relinked to one of its neighbors provided, again, that the global condition $f\leq 3$ remains satisfied.
One can easily see that this algorithm never creates loops in the tree.
Trees are then sampled according to~\eqref{eq:GranCanonicalDistribution} by implementing detailed balance through the classical Metropolis rule~\cite{metropolis}, for values $\beta\mu_3 = -1, 0, 1, 2$.
For each node density $\rho$, we verify that we have effectively reached the asymptotic large-$N$ regime by monitoring the mean branch-node fraction $\langle \phi_3\rangle$ as a function of $N$ (see Fig.~\ref{fig:2Ddata} in SM, notice also the exactness of Eq.~\eqref{eq:BranchingProbability} for ideal trees (black symbols)).
Plots of MF Eq.~\eqref{eq:BranchingProbability} (lines) and asymptotic MC results (symbols) as a function of $\beta\mu_3$, for node density $\rho$ and spatial dimension $d$ are shown in Fig.~\ref{fig:<n3>/N}, with the insets reporting the ratios between MC and MF values.
We notice that the agreement improves as a function of $\beta\mu_3$, $\rho$ and $d$, increasing from a good $\approx 78\%$ in $d=2$, $\rho=0$, $\beta\mu_3=-1$ to an excellent $\approx 95\%$ in $d=4$, $\rho=1$, $\beta\mu_3=2$.\footnote{These findings are in agreement with the known fact~\cite{Georges1991} that MF solutions of lattice models become more and more accurate with the increasing of space dimensionality. Moreover, by increasing $\rho$ we expect~\cite{DeGennesBook} partial screening of excluded-volume effects and, therefore, a better agreement between simulations and the MF solution, as also seen clearly in our data.}

{\it Conclusions} --
In this work, we have presented a field-theoretic method to study the thermodynamics of single, self-avoiding trees on a generic, regular lattice.
We have obtained a compact mean-field expression (Eq.~\eqref{eq:MFFreeEnergy}) for the SAT entropy that is, at the same time, significantly simpler and more general than others obtained in seemingly MF contexts~\cite{LubenskyIsaacson1979,Lubensky1981,deLosRios2000,Safran2002,Wagner2015}, and whose accuracy (see Table~\ref{tab:ComparisonExactMeanField} in SM) can in principle~\cite{BawendiFreed1988} be improved by the systematic inclusion of higher order terms beyond the saddle-point approximation utilized here.
Notably, our result provides a simple, analytical formula that is able to interpolate between dilute trees ($\rho \ll 1$) and spanning trees ($\rho=1$), as well as between linear and highly branched structures.
The results for all these different scenarios have been compared with exact results from the literature, highlighting the good accuracy of our MF approximation.
We have also provided predictions for the branching probability (Eq.~\eqref{eq:BranchingProbability}) that are in good agreement with computational results from Monte Carlo simulations.
In this regard, a new simple, yet effective algorithm for the simulation of spanning trees has also been presented.
Finally, with little additional effort the model can be extended to include the contribution of nearest-neighbor attractive interactions between tree nodes to model the situation of branched polymers in poor-solvent conditions~\cite{RubinsteinColby}.
In particular (see Sec.~\ref{sec:PoorSolventConditions} in SM), poor-solvent interactions do not affect the mean values of branch-node fractions found for simple excluded-volume conditions; at the same time, the expected globule-coil transition as a function of temperature is of the same kind of the one for simple linear ({\it i.e.} unbranched) polymers in agreement with works~\cite{Nemirovsky1992,deLosRios2000}.

{\it Acknowledgments} --
DM and AR acknowledge P. H. W. van der Hoek and R. Everaers for insightful discussions.
AR acknowledges financial support from PNRR Grant CN\_00000013\_CN-HPC, M4C2I1.4, spoke 7, funded by Next Generation EU.
AG acknowledges financial support by MIUR PRIN-COFIN2022 grant 2022JWAF7Y.

\section*{Data availability}
The data that support the findings of this study are available from the corresponding author upon reasonable request.

\bibliography{biblio}

\widetext
\clearpage
\begin{center}
\textbf{\Large Supplementary Material \\ \vspace*{1.5mm} Entropy of self-avoiding branching polymers: Mean-field theory and Monte Carlo simulations} \\
\vspace*{5mm}
Davide Marcato, Achille Giacometti, Amos Maritan, Angelo Rosa
\vspace*{10mm}
\end{center}

\setcounter{equation}{0}
\setcounter{figure}{0}
\setcounter{table}{0}
\setcounter{page}{1}
\setcounter{section}{0}
\setcounter{page}{1}
\makeatletter
\renewcommand{\theequation}{S\arabic{equation}}
\renewcommand{\thefigure}{S\arabic{figure}}
\renewcommand{\thetable}{S\arabic{table}}
\renewcommand{\thesection}{S\arabic{section}}
\renewcommand{\thepage}{S\arabic{page}}

\tableofcontents

\clearpage

\section{Relation between entropies of undirected and rooted-directed trees}\label{sec:UndirectedDirectedTrees}
In this Section, we discuss the quantitative relation between the entropies of undirected and rooted-directed SAT's at the basis of Eq.~\eqref{eq:DirectedUndirectedRelation} in main text.
For the sake of simplicity, we work with examples relative to the $2d$ square lattice; it is then quite straightforward to extend the arguments to more general lattices.

\begin{figure}
\centering
$$
\begin{array}{c}
\includegraphics[width=0.7\textwidth]{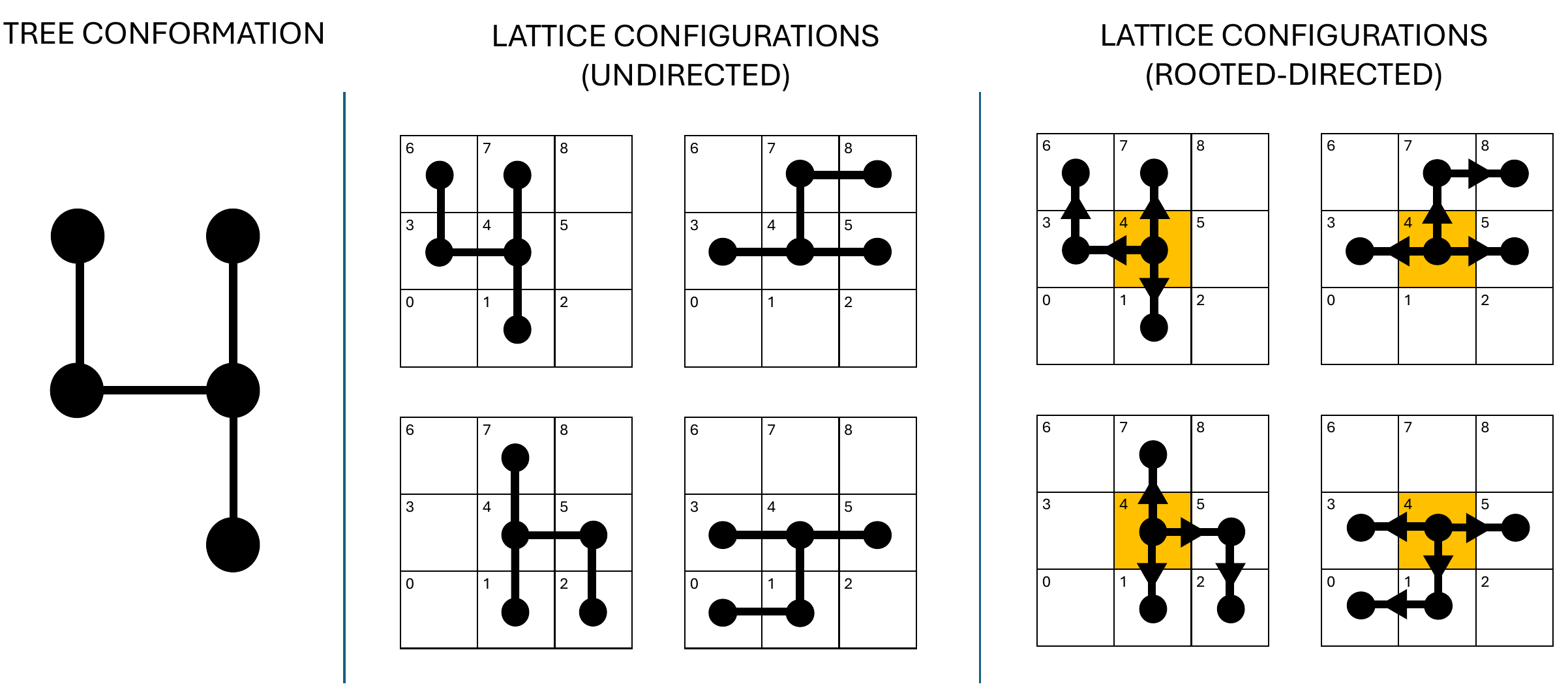} \\
\includegraphics[width=0.7\textwidth]{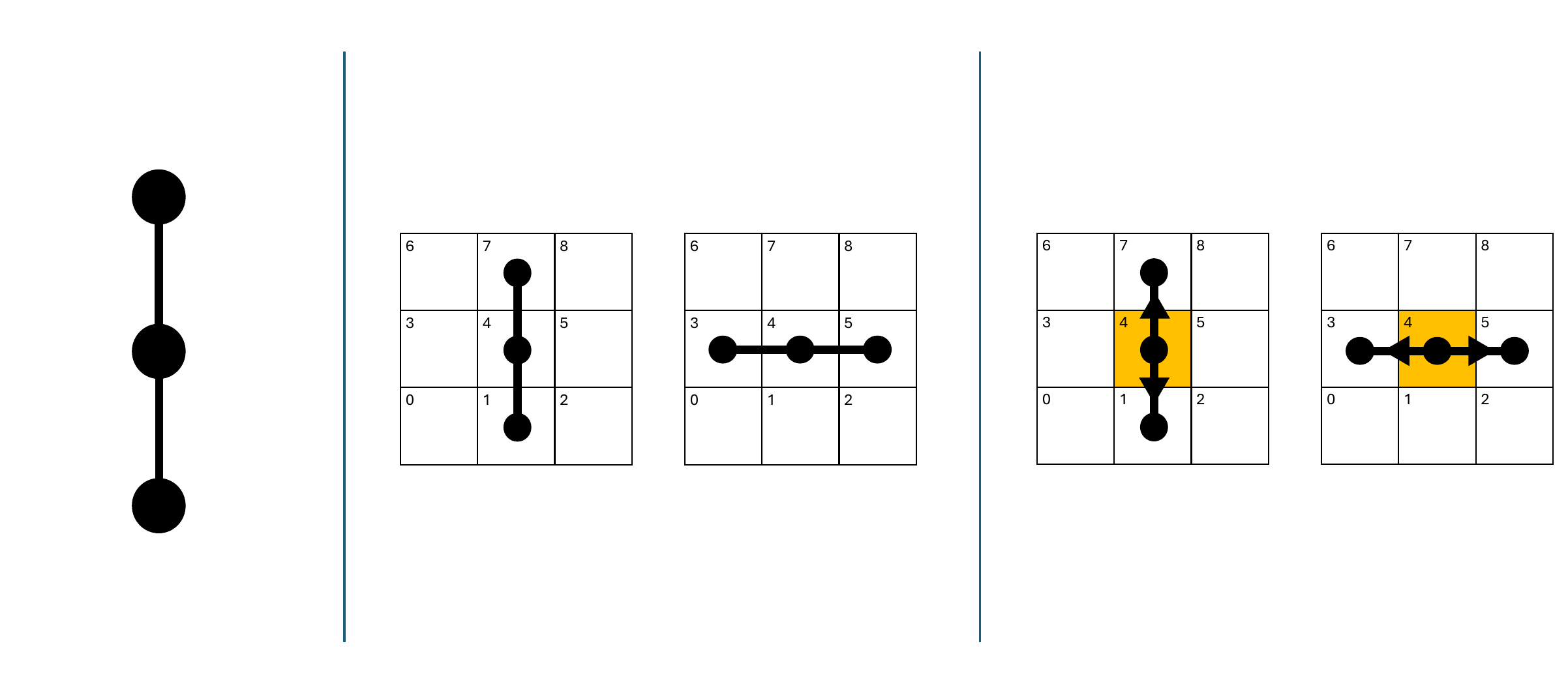} \\
\includegraphics[width=0.7\textwidth]{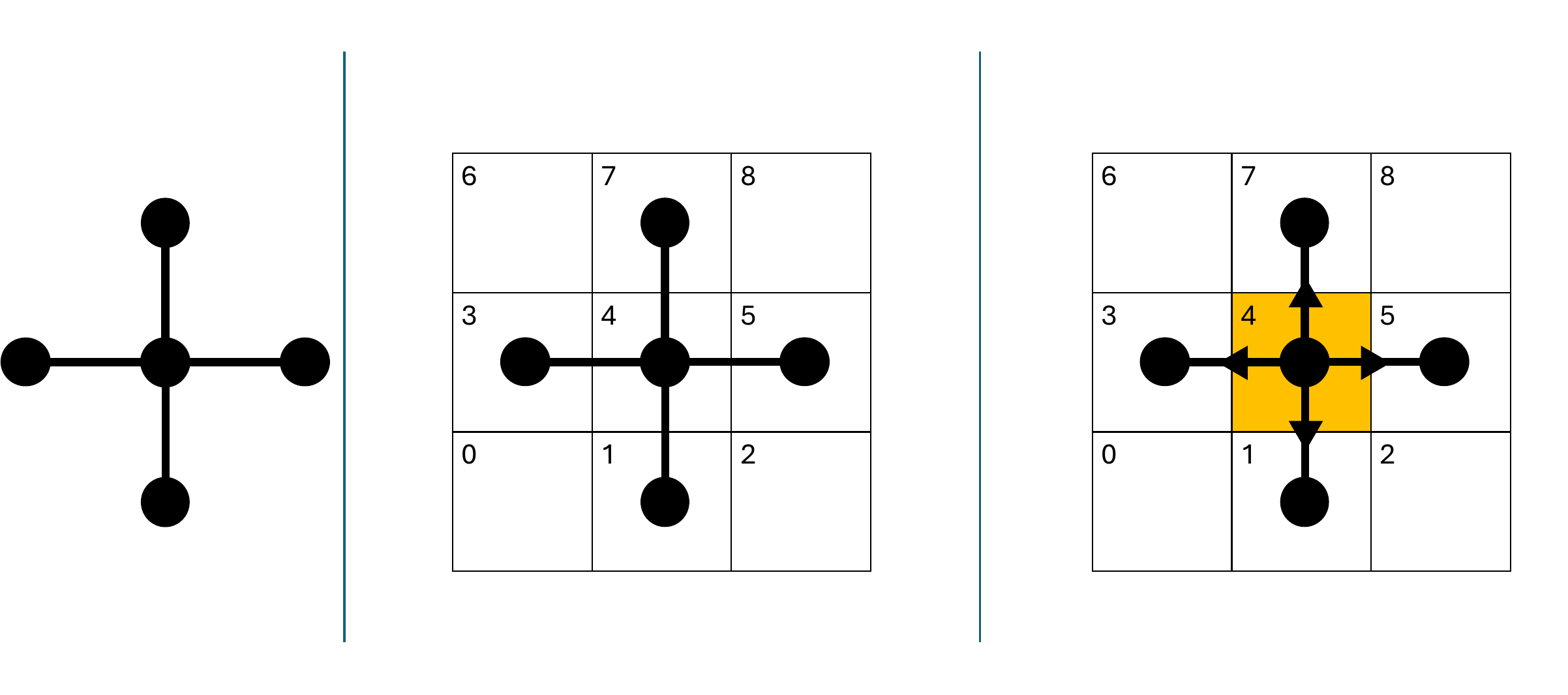}
\end{array}
$$
\caption{
Tree conformations (left column) and their corresponding lattice configurations as undirected trees (central column) and rooted-directed trees (right column) on the $3\times 3$ square lattice.
}
\label{fig:TreeConformationsUNdirectedDirected}
\end{figure}

To fix the ideas, consider the tree conformations shown in Fig.~\ref{fig:TreeConformationsUNdirectedDirected}.
\begin{itemize}
\item
(Top panels)
The tree conformation does not display rotational symmetry: as a consequence, there are $4$ different lattice configurations of undirected trees corresponding to the same conformation.
Then, each of the 4 configurations in the middle column can be translated to any of the sites of the lattice (which is possible because periodic boundary conditions are assumed).
Therefore, there are in total $4 \times 9 = 36$ undirected tree lattice configurations corresponding to the same polymer conformation.
For what concerns rooted-directed trees, the spatial position of the root has been {\it arbitrarily} assigned to the central lattice site.
It is then clear that, {\it at fixed root position}, the degeneracy of lattice configurations of rooted-directed trees due to rotations is still equal to $4$.
However, since we have fixed the position of the root, we have actually reduced the degeneracy due to translations: in fact, the tree can be translated as long as there is always a monomer at the site of the root.
Hence, each of the $4$ configurations depicted in the right column has a further translational degeneracy equal to the number of monomers: hence, the total degeneracy in this case is equal to $4\times 5 = 20$.
\item
(Middle panels)
Let us now consider the second polymer conformation, which displays partial rotational symmetry.
In the undirected case, the degeneracy of lattice configurations due to rotation is now $2$: the degeneracy due to translation remains instead equal to the number of lattice sites, {\it i.e.} $9$ in our example.
Again, for the rooted-directed lattice configurations the rotational degeneracy is the same as for the undirected case, while the translational degeneracy is equal to the number of monomers, {\it i.e.} $3$.
\item
(Bottom panels)
The last conformation is fully symmetric under rotations on the square lattice, therefore there is no degeneracy due to rotations for the lattice configurations, neither in the undirected nor in the rooted-directed case.
However, the degeneracy due to translations remains equal to $9$ for the undirected case and equal to $5$ for the rooted-directed case ({\it i.e.} the number of sites and the number of monomers, respectively).
\end{itemize}

By these examples, it should be clear that the degeneracy of each polymer conformation is {\it proportional} to $V$ for the undirected case and to $N$ for the rooted-directed case, with the proportionality constant depending on the symmetry of the conformations: however, {\it at fixed position of the root}, the proportionality constant is the same in both cases.
In other words, we have that 
\begin{equation}\label{eq:OmegaN_over_OmegaNx0}
\frac{\Omega_N}{\Omega_N^{{\mathbf x}_0}} = \frac{V}N \, .
\end{equation}
By taking the logarithm of both sides of Eq.~\eqref{eq:OmegaN_over_OmegaNx0} and dividing by $V$, Eq.~\eqref{eq:DirectedUndirectedRelation} in main text follows easily.

\section{Field-theoretic partition function and mean-field approximation}\label{sec:FieldTheoryCompleteDerivation}
In this Section, we illustrate in detail the derivation of the mean-field expression of the entropy (Eq.~\eqref{eq:MFFreeEnergy} in main text) for self-avoiding trees (SAT's).

\subsection{Spin $n$-vector model for SAT's and $n\to 0$ limit}\label{sec:On->0model}

\subsubsection{Formulation of the problem}\label{sec:On->0model-i}
We consider a regular lattice with coordination number (equal to the number of nearest neighbors per lattice site) $=q$, total number of sites $=V$ and with periodic boundary conditions; lengths are expressed in units of the lattice step length.
As explained in main text, the goal is to estimate $\Omega_N^{{\mathbf x}_0}(\left\{ N_f \right\})$, the total number of {\it rooted-directed} trees with root on the lattice site ${\mathbf x}_0$, $N$ nodes and $\left\{ N_f \right\} \equiv (N_3, ..., N_q)$ branch-nodes of functionality $f$.

For mathematical convenience, we start working in the ensemble in which the total number of bonds $N_b$ and the number of $f$-functional branch-nodes $N_f$ are not fixed, rather they are controlled by appropriate fugacities.
Then, the {\it grand canonical} partition function, $\Xi=\Xi(\kappa_b, \{ \kappa_f \})$,  of the system is given by
\begin{equation}\label{eq:GrandCanonicalZ}
\Xi 
= \sum_{\{ {\mathcal C} \}_{{\mathbf x}_0}} \bigg( \kappa^{N_b({\mathcal C})}_b \, \prod_{f=3}^q \kappa_f^{N_f({\mathcal C})} \bigg) \, ,
\end{equation}
where $\{ {\mathcal C} \}_{{\mathbf x}_0}$ is the set of all possible conformations of rooted-directed lattice tree rooted on site ${\mathbf x}_0$.
The total number of bonds in a given conformation ${\mathcal C}$ is indicated with $N_b({\mathcal C})$ and $\kappa_b$ is the corresponding bond fugacity; then, $N_f({\mathcal C})$ is the number of $f$-functional branch-nodes and $\kappa_f$ is the corresponding fugacity ($=e^{\beta \mu_f}$, where $\mu_f$ is the chemical potential for the $f$-functional branch-nodes, see Eq.~(8) 
in main text).

Assign now two $n$-component spin vectors, ${\mathbf S}({\mathbf x}) \equiv ( S_1({\mathbf x}), ..., S_n({\mathbf x}) )$ and $\bar{{\mathbf S}}({\mathbf x}) \equiv ( {\bar S}_1({\mathbf x}), ..., {\bar S}_n({\mathbf x}) )$, to each lattice site of spatial position ${\mathbf x}$, and define the {\it trace operation} (indicated by $\langle \cdot \rangle$) on the spin vectors characterized by the following mathematical properties:
\begin{itemize}
\item[(I)]
Spin vectors on different sites are independent of each other.
\item[(II)]
$\langle 1 \rangle = 1$.
\item[(III)]
$\langle S_i({\mathbf x}) \rangle = 1$, for $i = 1, ..., n$.
\item[(IV)]
$\langle S_i({\mathbf x}) \, S_j({\mathbf x}) \text{(anything)} \rangle = 0$, for every $i,j$.
\item[(V)]
$\left\langle \prod_{i=1}^k {\bar S}_i({\mathbf x}) \right\rangle = 0$, for every $k \geq 1$.
\item[(VI)]
$\langle S_i({\mathbf x}) \, {\bar S}_{j_1}({\mathbf x}) \, {\bar S}_{j_2}({\mathbf x}) \, \dots \, {\bar S}_{j_k}({\mathbf x}) \rangle =
\begin{cases}
\delta_{i,j_1} & \text{if $k=1$}; \\
\delta_{i,j_1,j_2} \, [\kappa_3 \, (1-\delta_{{\mathbf x}, {\mathbf x}_0}) + \delta_{{\mathbf x}, {\mathbf x}_0}] & \text{if $k=2$}; \\
\delta_{i,j_1,j_2,j_3} \, [\kappa_4 \, (1-\delta_{{\mathbf x},{\mathbf x}_0}) + \kappa_3 \, \delta_{{\mathbf x},{\mathbf x}_0}] & \text{if $k=3$}; \\
\vdots\\
\delta_{i,j_1,j_2, \dots, j_{q-1}} \, [\kappa_q \, (1-\delta_{{\mathbf x},{\mathbf x}_0}) + \kappa_{q-1} \, \delta_{{\mathbf x},{\mathbf x}_0}] & \text{if $k = q-1$}; \\
\delta_{i,j_1,j_2, \dots, j_q} \, \kappa_q \, \delta_{{\mathbf x},{\mathbf x}_0} & \text{if $k = q$}; \\
0 & \text{otherwise.}
\end{cases}
$,\\
for every $\mathbf x$ and ${\mathbf x}'$ and where the generalized Kronecker's symbol $\delta_{ij...}$ is $=1$ for identical subscripts and $=0$ otherwise.
\end{itemize}
Given these properties, the following relation holds
\begin{equation}\label{eq:ZDiagramaticExpansion}
\Xi 
= \lim_{n \to 0} \, \bigg\langle
S_1({\mathbf x}_0)
\prod_{{\mathbf x}, {\mathbf x}'} \bigg( 1 + \kappa_b \, \bar{\mathbf S}({\mathbf x}) \cdot {\mathbf S}({\mathbf x}') \, \Delta({\mathbf x}, {\mathbf x}') \bigg) 
\bigg\rangle
\end{equation}
where $\Delta(\bold{x}, \bold{x}')$ is the adjacency matrix of the lattice.
Then, the term ``$1$'' corresponds to an empty lattice site, the term ``$S_1({\mathbf x}_0)$'' corresponds to a node in the lattice position of the root ${\mathbf x}_0$, whereas the term ``$\bar{\mathbf S}({\mathbf x}) \cdot {\mathbf S}({\mathbf x}')$'' represents a {\it directed} bond from a node on the site ${\mathbf x}$ to a node on the site ${\mathbf x}'$.

\begin{figure}
$$
\begin{array}{ccc}
\includegraphics[width=0.28\textwidth]{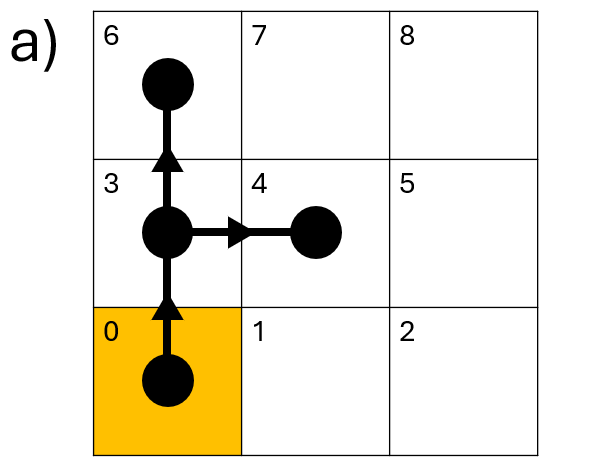} & \includegraphics[width=0.28\textwidth]{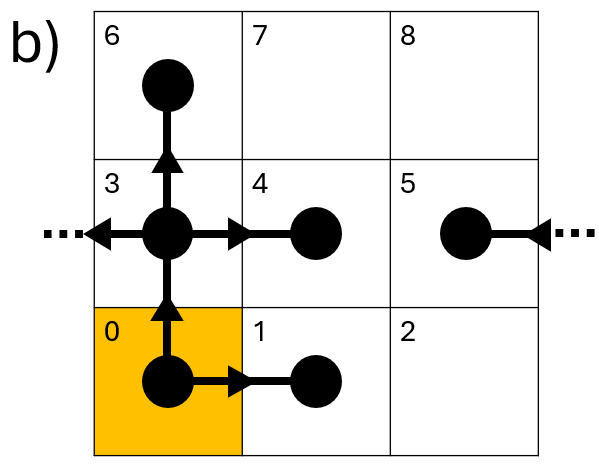} & \includegraphics[width=0.28\textwidth]{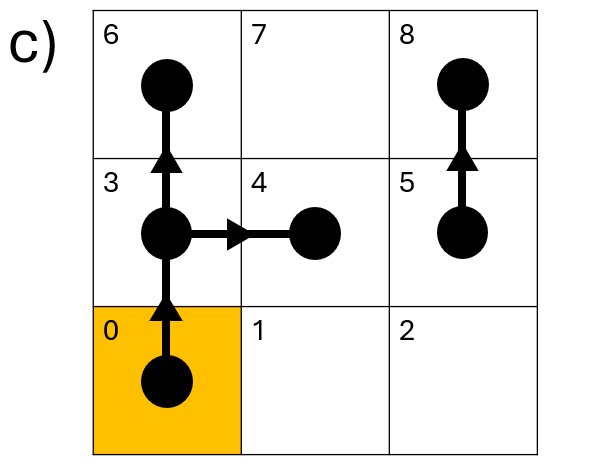}
\end{array}
$$
$$
\begin{array}{cc}
\includegraphics[width=0.28\textwidth]{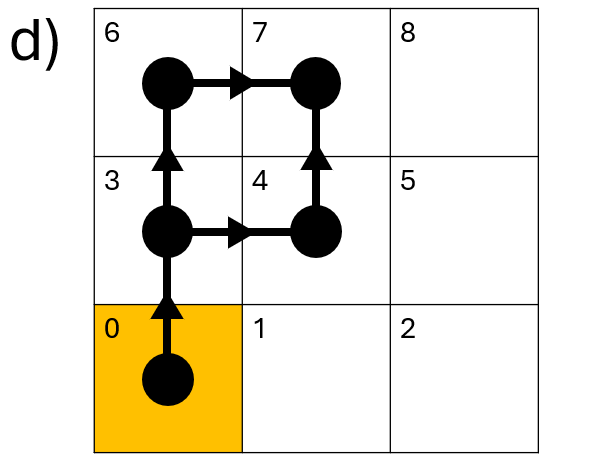} &
\includegraphics[width=0.28\textwidth]{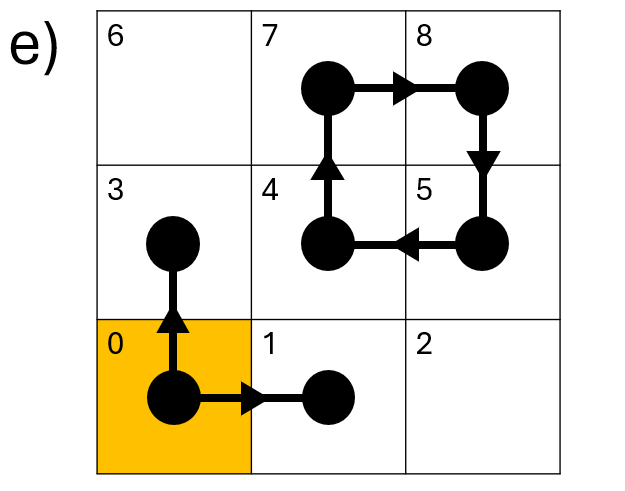}
\end{array}
$$
\caption{
Self-avoiding tree conformations on the $3\times3$ square lattice with periodic boundary conditions representing some typical terms of the diagrammatic expansion of Eq.~\eqref{eq:ZDiagramaticExpansion}.
Lattice sites are numbered from $0$ to $8$, the spatial position of the tree root is ${\mathbf x}_0$ (bottom left corner, in yellow).
}
\label{fig:IllustrativeConfigurations}
\end{figure}

\subsubsection{Validation of Eq.~\eqref{eq:ZDiagramaticExpansion}}\label{sec:On->0model-ii}
To show how by these ingredients one can construct the correct partition function~\eqref{eq:ZDiagramaticExpansion} of the system, consider the example conformations in Fig.~\ref{fig:IllustrativeConfigurations}.

\noindent
Fig.~\ref{fig:IllustrativeConfigurations}(a) --
This conformation is represented by the term:
\begin{equation}\label{eq:TreeExample1-a}
\kappa_b^3 \,
\langle
S_1({\mathbf x}_0)
(\bar{\mathbf S}({\mathbf x}_0) \cdot {\mathbf S}({\mathbf x}_3))
(\bar{\mathbf S}({\mathbf x}_3) \cdot {\mathbf S}({\mathbf x}_4))
(\bar{\mathbf S}({\mathbf x}_3) \cdot {\mathbf S}({\mathbf x}_6))
\rangle \, ,
\end{equation}
which is equivalent to
\begin{equation}\label{eq:TreeExample1-b}
\kappa_b^3 \,
\sum_{j_1} \sum_{j_2} \sum_{j_3}
\langle
S_1({\mathbf x}_0) \,
\bar{S}_{j_1}({\mathbf x}_0) S_{j_1}({\mathbf x}_3) \, 
\bar{S}_{j_2}({\mathbf x}_3) S_{j_2}({\mathbf x}_4) \, 
\bar{S}_{j_3}({\mathbf x}_3) S_{j_3}({\mathbf x}_6)
\rangle \, .
\end{equation}
By exploiting trace property (I) (spin vectors on different sites are independent), Eq.~\eqref{eq:TreeExample1-b} becomes:
\begin{equation}\label{eq:TreeExample1-c}
\kappa_b^3 \,
\sum_{j_1} \sum_{j_2} \sum_{j_3} \,
\langle S_1({\mathbf x}_0) \bar{S}_{j_1}({\mathbf x}_0) \rangle \,
\langle S_{j_1}({\mathbf x}_3) \bar{S}_{j_2}({\mathbf x}_3) \bar{S}_{j_3}({\mathbf x}_3) \rangle \,
\langle S_{j_2}({\mathbf x}_4) \rangle \,
\langle S_{j_3}({\mathbf x}_6) \rangle \, ,
\end{equation}
which, by the rest of the properties (II)-(VI), simplifies to
\begin{equation}\label{eq:TreeExample1-d}
\kappa_b^3 \, \kappa_3 \,
\sum_{j_1}\sum_{j_2}\sum_{j_3} \,
\delta_{1,j_1} \, \delta_{j_1,j_2,j_3} = \kappa_b^3 \, \kappa_3 \, .
\end{equation}
%

\noindent
Fig.~\ref{fig:IllustrativeConfigurations}(b) --
By the same procedure as before, one gets
\begin{equation}\label{eq:TreeExample1-e}
\kappa_b^5
\sum_{\substack{j_1, j_2, j_3 \\ j_4, j_5}}
\langle S_1({\mathbf x}_0) \bar{S}_{j_1}(\bold{x}_0) \bar{S}_{j_2}({\mathbf x}_0)\rangle \,
\langle S_{j_2}({\mathbf x}_3) \bar{S}_{j_3}({\mathbf x}_3) \bar{S}_{j_4}({\mathbf x}_3) \bar{S}_{j_5}({\mathbf x}_3) \rangle \,
\langle S_{j_1}({\mathbf x}_1) \rangle \,
\langle S_{j_3}({\mathbf x}_4) \rangle \,
\langle S_{j_4}({\mathbf x}_5) \rangle \,
\langle S_{j_5}({\mathbf x}_6) \rangle \, ,
\end{equation}
which leads to:
\begin{equation}\label{eq:TreeExample1-f}
\kappa_b^5 \, \kappa_4
\sum_{\substack{j_1, j_2, j_3 \\ j_4, j_5}}
\delta_{1,j_1,j_2} \, \delta_{j_2,j_3,j_4,j_5}
= \kappa_b^5 \, \kappa_4 \, .
\end{equation}
%

\noindent
Fig.~\ref{fig:IllustrativeConfigurations}(c) --
Here we show how the trace properties rule out almost every (see later) {\it disconnected} conformation from the partition function.
In fact, the one in the figure corresponds to the term:
\begin{equation}\label{eq:TreeExample1-g}
\kappa_b^4 \,
\langle
S_1({\mathbf x}_0) \,
(\bar{{\mathbf S}}({\mathbf x}_0) \cdot {\mathbf S}({\mathbf x}_3)) \,
(\bar{\mathbf S}({\mathbf x}_3) \cdot {\mathbf S}({\mathbf x}_4)) \,
(\bar{\mathbf S}({\mathbf x}_3) \cdot {\mathbf S}({\mathbf x}_6)) \,
(\bar{\mathbf S}({\mathbf x}_5) \cdot {\mathbf S}({\mathbf x}_8))
\rangle \, .
\end{equation}
It is easy to see that, after factorization, the expression~\eqref{eq:TreeExample1-g} contains a factor $\sim \langle \bar{S}_j({\mathbf x}_5) \rangle$ which, by the trace property (V), is $=0$ and so the weight of the entire conformation.

\noindent
Fig.~\ref{fig:IllustrativeConfigurations}(d) --
Similarly, the trace properties rule out any conformation that contains a {\it loop}.
For the conformation in the figure, we have the term:
\begin{equation}\label{eq:TreeExample1-h}
\kappa_b^5 \,
\langle
S_1({\mathbf x}_0) \,
(\bar{\mathbf S}({\mathbf x}_0) \cdot {\mathbf S}({\mathbf x}_3)) \,
(\bar{\mathbf S}({\mathbf x}_3) \cdot {\mathbf S}({\mathbf x}_4)) \,
(\bar{\mathbf S}({\mathbf x}_3) \cdot {\mathbf S}({\mathbf x}_6)) \,
(\bar{\mathbf S}({\mathbf x}_6) \cdot {\mathbf S}({\mathbf x}_7)) \,
(\bar{\mathbf S}({\mathbf x}_4) \cdot {\mathbf S}({\mathbf x}_7))
\rangle \, .
\end{equation}
Again, after factorization, it is straightforward to see that Eq.~\eqref{eq:TreeExample1-h} contains a factor $\sim \langle S_{j_1}({\mathbf x}_7) S_{j_2}({\mathbf x}_7) \rangle$ that is $=0$ because of the trace property (IV).

\noindent
Fig.~\ref{fig:IllustrativeConfigurations}(e) --
The latest possible situation, and probably the trickiest, is the one shown in the last panel.
It corresponds to the term:
\begin{equation}\label{eq:TreeExample1-i}
\kappa_b^6 \,
\langle S_1({\mathbf x}_0) \,
(\bar{\mathbf S}({\mathbf x}_0) \cdot {\mathbf S}({\mathbf x}_1)) \,
(\bar{\mathbf S}({\mathbf x}_0) \cdot {\mathbf S}({\mathbf x}_3)) \,
(\bar{\mathbf S}({\mathbf x}_4) \cdot {\mathbf S}({\mathbf x}_7)) \,
(\bar{\mathbf S}({\mathbf x}_7) \cdot {\mathbf S}({\mathbf x}_8)) \,
(\bar{\mathbf S}({\mathbf x}_8) \cdot {\mathbf S}({\mathbf x}_5)) \,
(\bar{\mathbf S}({\mathbf x}_5) \cdot {\mathbf S}({\mathbf x}_4))
\rangle \, ,
\end{equation}
and, as it can be easily verified, is $\neq 0$ even though it corresponds to a prohibited conformation.
In fact, Eq.~\eqref{eq:TreeExample1-i} is just equal to:
\begin{equation}\label{eq:TreeExample1-l}
\kappa_b^6 \,
\sum_{\substack{j_1,j_2 \\ j_3,j_4 \\ j_5, j_6}} \,
\langle S_1({\mathbf x}_0) \bar{S}_{j_1}({\mathbf x}_0) \bar{S}_{j_2}({\mathbf x}_0) \rangle \,
\langle S_{j_1}({\mathbf x}_1) \rangle \,
\langle S_{j_2}({\mathbf x}_3) \rangle \,
\langle \bar{S}_{j_3}({\mathbf x}_4) S_{j_6}({\mathbf x}_4) \rangle \,
\langle \bar{S}_{j_6}({\mathbf x}_5) S_{j_5}({\mathbf  x}_5) \rangle \,
\langle \bar{S}_{j_4}(\bold{x}_7)S_{j_3}({\mathbf x}_7) \rangle \,
\langle \bar{S}_{j_5}({\mathbf x}_8) S_{j_4}({\mathbf x}_8) \rangle \, ,
\end{equation}
which, based on the trace properties, simplifies to:
\begin{equation}
\kappa_b^6 \,
\sum_{\substack{j_1,j_2}} \,
\delta_{1,j_1,j_2} \, 
\sum_{\substack{j_3,j_4 \\ j_5,j_6}} \,
\delta_{j_3,j_4} \delta_{j_4,j_5} \delta_{j_5,j_6} \delta_{j_6,j_3} = \kappa_b^6 \, n \, .
\end{equation}
The weight of the conformation is thus proportional to $n$ yet, owing to the limit $n \to 0$ (see Eq.~\eqref{eq:ZDiagramaticExpansion}), these conformations are also removed from the partition function.
In fact, it is not difficult to realize that the $n \to 0$ limit helps precisely to get rid of the unwanted case of isolated loops.

To summarize, we have shown that Eq.~\eqref{eq:ZDiagramaticExpansion} effectively counts {\it only} rooted-directed SAT conformations and, importantly, with the expected statistical weight given by the fugacities $\kappa_b$ and $\{ \kappa_f \}$.
We can now proceed with the derivation of the field-theoretic form of Eq.~\eqref{eq:ZDiagramaticExpansion}.

\subsubsection{Field-theoretic form of Eq.~\eqref{eq:ZDiagramaticExpansion}}\label{sec:On->0model-iii}
According to the trace properties (I)-(VI) and taking into account the discussion of Sec.~\ref{sec:On->0model-ii}, Eq.~\eqref{eq:ZDiagramaticExpansion} can be written as the following:
\begin{equation}\label{eq:ZDiagramaticExpansion-exp}
\Xi 
= \lim_{n \to 0}
\bigg\langle
S_1({\mathbf x}_0)
\prod_{{\mathbf x}, {\mathbf x}'}
\exp\bigg[ \kappa_b \bar{\mathbf S}({\mathbf x}) \cdot {\mathbf S}({\mathbf x}') \, \Delta({\mathbf x}, {\mathbf x}') \bigg] \bigg\rangle =
\lim_{n \to 0} \bigg\langle S_1({\mathbf x}_0) \exp\bigg[ \kappa_b \sum_{{\mathbf x}, {\mathbf x}'} \bar{\mathbf S}({\mathbf x}) \cdot {\mathbf S}({\mathbf x}') \, \Delta({\mathbf x}, {\mathbf x}') \bigg] \bigg\rangle \, ,
\end{equation}
where the first equality is justified by the fact that in the expansion of the exponential all terms beyond first order have weight $=0$ (see Sec.~\ref{sec:On->0model-ii} for an explanation of this point). 
In order to obtain the field-theoretic form of Eq.~\eqref{eq:ZDiagramaticExpansion-exp}, we apply a generalization of the standard Hubbard-Stratonovich formula~\cite{ZinnJustin2021} to it and get:
\begin{multline}\label{eq:ZDiagramaticExpansion-Factor}
\exp\bigg[ \kappa_b \sum_{{\mathbf x}, {\mathbf x}'} \bar{\mathbf S}({\mathbf x}) \cdot {\mathbf S}({\mathbf x}') \, \Delta({\mathbf x}, {\mathbf x}') \bigg] =
\prod_{\alpha=1}^n \exp\bigg[ \kappa_b \sum_{{\mathbf x}, {\mathbf x}'} \bar{S}_\alpha({\mathbf x}) S_\alpha({\mathbf x}') \, \Delta({\mathbf x}, {\mathbf x}') \bigg] = \\
\prod_{\alpha=1}^n (\det \Delta^{-1}) \int\prod_{\mathbf x} \frac{d\varphi_\alpha({\mathbf x}) \, d\bar{\varphi}_{\alpha}({\mathbf x})}{2\pi i} \, \exp\bigg[ -\sum_{{\mathbf x}, {\mathbf x}'} \bar{\varphi}_{\alpha}({\mathbf x}) \, \Delta^{-1}({\mathbf x}, {\mathbf x}') \, \varphi_\alpha({\mathbf x}') + \sqrt{\kappa_b} \, \sum_{{\mathbf x}} \bigg( \bar{S}_\alpha({\mathbf x}) \varphi_\alpha({\mathbf x}) + S_{\alpha}({\mathbf x}) \bar{\varphi}_\alpha({\mathbf x}) \bigg) \bigg] = \\
(\det \Delta^{-1})^n \int \prod_{\mathbf x} \frac{d\boldsymbol{\varphi}({\mathbf x}) \, d\bar{\boldsymbol{\varphi}}({\mathbf x})}{(2\pi i)^n} \exp\bigg[ -\sum_{{\mathbf x}, {\mathbf x}'} \bar{\boldsymbol{\varphi}}({\mathbf x}) \cdot \boldsymbol{\varphi}({\mathbf x}') \, \Delta^{-1}({\mathbf x}, {\mathbf x}') + \sqrt{\kappa_b} \, \sum_{\mathbf x} \bigg( \bar{\mathbf S}({\mathbf x}) \cdot \boldsymbol{\varphi}({\mathbf x}) + {\mathbf S}(\bold{x}) \cdot \bar{\boldsymbol{\varphi}}({\mathbf x}) \bigg) \bigg] \, ,
\end{multline}
where $\boldsymbol{\varphi}({\mathbf x})$ and $\bar{\boldsymbol{\varphi}}({\mathbf x})$ are $n$-component vector fields relative to the lattice site ${\mathbf x}$.
By plugging Eq.~\eqref{eq:ZDiagramaticExpansion-Factor} back into the expression for $\Xi$ 
we get
\begin{eqnarray}
\Xi 
= \lim_{n \to 0} \int \prod_{\mathbf x} d\boldsymbol{\varphi}({\mathbf x}) \, d\bar{\boldsymbol{\varphi}}({\mathbf x}) \exp\bigg[ -\sum_{{\mathbf x}, {\mathbf x}'} \bar{\boldsymbol{\varphi}}({\mathbf x}) \cdot \boldsymbol{\varphi}({\mathbf x}') \, \Delta^{-1}({\mathbf x}, {\mathbf x}') \bigg] \bigg\langle S_1({\mathbf x}_0) \exp\bigg[ \sqrt{\kappa_b} \, \sum_{\mathbf x} \bigg( \bar{\mathbf S}({\mathbf x}) \cdot \boldsymbol{\varphi}({\mathbf x}) + {\mathbf S}({\mathbf x}) \cdot \bar{\boldsymbol{\varphi}}({\mathbf x}) \bigg) \bigg] \bigg\rangle \, , \nonumber\\
\end{eqnarray}
where we have already taken the limit for $(\det\Delta^{-1})^n \to 1$ and $(2\pi i)^n \to 1$.
At this point, the term inside the $\langle \rangle$ can be factorized, making the calculation of the trace much easier.
By expanding again the exponential and using the trace properties (I)-(VI), one finally gets the expression:
\begin{eqnarray}\label{eq:FieldTheoreticZ}
\Xi 
& = &
\lim_{n \to 0} \int\prod_{\mathbf x} d\boldsymbol{\varphi}(\bold{x}) \, d\bar{\boldsymbol{\varphi}}({\mathbf x}) \exp\bigg[ -\sum_{{\mathbf x}, {\mathbf x}'} \, \Delta^{-1}({\mathbf x}, {\mathbf x}') \, \bar{\boldsymbol{\varphi}}({\mathbf x}) \cdot \boldsymbol{\varphi}({\mathbf x}') \nonumber\\
& & + \log\bigg( 1+ \sqrt{\kappa_b} \, \varphi_1({\mathbf x}_0) + \frac{\kappa_b}2 \, \varphi_1({\mathbf x}_0)^2 + \sum_{f=3}^q \kappa^{f/2}_b \, \kappa_f \, \frac{\varphi_1^f({\mathbf x}_0)}{f!} \bigg) \nonumber\\
& & + \sum_{{\mathbf x} \neq {\mathbf x}_0} \log \bigg( 1 + \sqrt{\kappa_b} \, \sum_{\alpha=1}^n \bar{\varphi}_{\alpha}({\mathbf x}) + \kappa_b \sum_{\alpha=1}^n \bar{\varphi}_{\alpha}({\mathbf x}) \, \varphi_{\alpha}({\mathbf x}) + \sum_{\alpha=1}^n \bar{\varphi}_{\alpha}({\mathbf x}) \sum_{f=3}^q \kappa^{f/2}_b \, \kappa_f \, \frac{\varphi_{\alpha}^{f-1}({\mathbf x})}{(f-1)!} \bigg) \bigg] \, .
\end{eqnarray}
Notice that Eq.~\eqref{eq:FieldTheoreticZ} is formally {\it exact}.

\subsubsection{Direct validation of Eq.~\eqref{eq:FieldTheoreticZ} by Wick's theorem}\label{sec:UseWickTheorem}
Eq.~\eqref{eq:FieldTheoreticZ} can be readily validated by means of (a classical version of) Wick's theorem~\cite{ZinnJustin2021}.
To show it, we take the logarithms and write Eq.~\eqref{eq:FieldTheoreticZ} as
\begin{eqnarray}\label{eq:FieldTheoreticZNoLog}
\Xi 
& = &
\lim_{n \to 0} \int\prod_{\mathbf x} d\boldsymbol{\varphi}(\bold{x}) \, d\bar{\boldsymbol{\varphi}}({\mathbf x}) \exp\bigg[ -\sum_{{\mathbf x}, {\mathbf x}'} \, \Delta^{-1}({\mathbf x}, {\mathbf x}') \, \bar{\boldsymbol{\varphi}}({\mathbf x}) \cdot \boldsymbol{\varphi}({\mathbf x}') \bigg]\nonumber\\
& & \times \bigg( 1+ \sqrt{\kappa_b} \, \varphi_1({\mathbf x}_0) + \frac{\kappa_b}2 \, \varphi_1({\mathbf x}_0)^2 + \sum_{f=3}^q \kappa^{f/2}_b \, \kappa_f \, \frac{\varphi_1^f({\mathbf x}_0)}{f!} \bigg) \nonumber\\
& & \times \prod_{{\mathbf x} \neq {\mathbf x}_0} \bigg( 1 + \sqrt{\kappa_b} \, \sum_{\alpha=1}^n \bar{\varphi}_{\alpha}({\mathbf x}) + \kappa_b \sum_{\alpha=1}^n \bar{\varphi}_{\alpha}({\mathbf x}) \, \varphi_{\alpha}({\mathbf x}) + \sum_{\alpha=1}^n \bar{\varphi}_{\alpha}({\mathbf x}) \sum_{f=3}^q \kappa^{f/2}_b \, \kappa_f \, \frac{\varphi_{\alpha}^{f-1}({\mathbf x})}{(f-1)!} \bigg)\, .
\end{eqnarray}
Then, computing the integrals of the various terms which appear in Eq.~\eqref{eq:FieldTheoreticZNoLog} is equivalent to determining the expectation values of polynomials for a multivariate Gaussian distribution with covariance matrix $\Delta^{-1}$.
In particular, we are making use of the following form of the theorem~\cite{ZinnJustin2021}:
\begin{equation}\label{eq:WickTheorem}
\langle \varphi_{\alpha_1}(\bold{x}_{i_1}) \bar{\varphi}_{\beta_1}(\bold{x}_{j_1}) \dots \varphi_{\alpha_k}(\bold{x}_{i_k})\bar{\varphi}_{\beta_k}(\bold{x}_{j_k}) \rangle_{\text{Gauss}} = \sum_{\substack{\text{all permutations $P$} \\\text{of $\{1, \dots,k\}$}}}\delta_{\alpha_1\beta_{P_1}}\Delta(\bold{x}_{i_1}, \bold{x}_{j_{P_1}}) \dots \delta_{\alpha_k\beta_{P_k}}\Delta(\bold{x}_{i_k}, \bold{x}_{j_{P_k}}) \, ,
\end{equation}
with the notation,
\begin{equation}\label{eq:MomentsGaussDistribution}
\langle \cdots \rangle_{\text{Gauss}}
\equiv
(2i\pi)^{Vn}(\det \Delta)^n\int\prod_{\mathbf x} d\boldsymbol{\varphi}(\bold{x}) \, d\bar{\boldsymbol{\varphi}}({\mathbf x}) ( \cdots )\exp\bigg[ -\sum_{{\mathbf x}, {\mathbf x}'} \, \Delta^{-1}({\mathbf x}, {\mathbf x}') \, \bar{\boldsymbol{\varphi}}({\mathbf x}) \cdot \boldsymbol{\varphi}({\mathbf x}') \bigg] \, ,
\end{equation}
and where all the moments with a non-equal number of $\varphi$ and $\bar{\varphi}$ terms give $0$.
For instance, with reference to Fig.~\ref{fig:IllustrativeConfigurations}, let us consider
\begin{multline}\label{eq:ExampleWickTheorem}
\bigg\langle \sqrt{\kappa_b} \, \varphi_1({\mathbf x}_0) \bigg(\sum_{\alpha_1}\bar{\varphi}_{\alpha_1}({\mathbf x}_3)\kappa^{3/2}_b\kappa_3\frac{\varphi^{2}_{\alpha_1}({\mathbf x}_3)}{2}\bigg) \bigg(\sqrt{\kappa_b}\sum_{\alpha_2}\bar{\varphi}_{\alpha_2}({\mathbf x}_4)\bigg) \bigg(\sqrt{\kappa_b}\sum_{\alpha_3}\bar{\varphi}_{\alpha_3}({\mathbf x}_6)\bigg) \bigg\rangle_{\text{Gauss}} = \\
\frac{\kappa^3_b\kappa_3}{2}\sum_{\alpha_1, \alpha_2, \alpha_3}\langle \varphi_1({\mathbf x}_0)\bar{\varphi}_{\alpha_1}({\mathbf x}_3)\varphi_{\alpha_1}({\mathbf x}_3)\bar{\varphi}_{\alpha_2}({\mathbf x}_4) \varphi_{\alpha_1}({\mathbf x}_3)\bar{\varphi}_{\alpha_3}({\mathbf x}_6) \rangle_{\text{Gauss}} \, .
\end{multline}
By using Eq.~\eqref{eq:WickTheorem}, it is easy to see that in the limit $n \to 0$ Eq.~\eqref{eq:ExampleWickTheorem} simplifies to:
\begin{equation}
\kappa^{3}_b \, \kappa_3 \, \Delta({\mathbf x}_0, {\mathbf x}_3)\Delta({\mathbf x}_3, {\mathbf x}_4) \Delta({\mathbf x}_3, {\mathbf x}_6) = \kappa^3_b \, \kappa_3 \, ,
\end{equation}
which corresponds (see Eq.~\eqref{eq:TreeExample1-d}) to the configuration depicted in panel (a) of Fig.~\ref{fig:IllustrativeConfigurations}.
By following a similar procedure for all the different terms in Eq.~\eqref{eq:FieldTheoreticZNoLog}, it can be checked that Eq.~\eqref{eq:FieldTheoreticZ} indeed counts all (and only) the SAT configurations with the correct weight based on the total number of bonds and of $f$-functional nodes.
In particular, a term $\varphi_{\alpha}({\mathbf x})$ indicates a bond exiting from site ${\mathbf x}$, whereas $\bar{\varphi}_{\alpha}({\mathbf x})$ indicates a bond entering in site ${\mathbf x}$.

\subsection{Saddle-point approximation}\label{sec:SaddlePointApprox}
Although exact, the integral in Eq.~\eqref{eq:FieldTheoreticZ} cannot be computed directly.
We are therefore looking for a uniform saddle-point (or, mean-field-like~\cite{Orland1985,Orland1996,Higuchi1998,Marcato2023,Marcato2024}) approximation.
First, we search for the stationary points of the integrand and then look for solutions of the saddle-point equations of the kind $\boldsymbol{\varphi}({\mathbf x}) = (\varphi, 0, \dots, 0)$ and $\bar{\boldsymbol{\varphi}}({\mathbf x}) = (\bar{\varphi}, 0, \dots, 0)$, $\forall \bold{x}$. 
This particular choice is justified by the explicit presence of the term $S_1$ that breaks the $n$-vector symmetry.
The approximate expression for the partition function then becomes:
\begin{equation}\label{eq:ApproximateZ}
\Xi 
\simeq \exp\bigg[ -V \frac{\varphi\bar{\varphi}}{q} + \log\bigg( 1 + \sqrt{\kappa_b} \, \varphi + \frac{\kappa_b}2 \varphi^2 + \sum_{f=3}^q \kappa_b^{f/2} \kappa_f \, \frac{\varphi^f}{f!} \bigg) \, + \, (V-1) \log\bigg( 1 + \sqrt{\kappa_b} \bar{\varphi} + \kappa_b \bar{\varphi}\varphi + \bar{\varphi} \sum_{f=3}^q \kappa_b^{f/2} \kappa_f \frac{\varphi^{f-1}}{(f-1)!} \bigg)\bigg] \, ,
\end{equation}
where $\varphi$ and $\bar{\varphi}$ are the solutions of the following saddle-point equations:
\begin{eqnarray}
\frac{\bar{\varphi}}{q}
& = & \frac1V \frac{\sqrt{\kappa_b} + \kappa_b\varphi + \sum_{f=3}^{q}\kappa^{f/2}_b\kappa_f \frac{\varphi^{f-1}}{(f-1)!}}{1 + \sqrt{\kappa_b}\varphi + \kappa_b\varphi^2 + \sum_{f=3}^{q}\kappa^{f/2}_b\kappa_f \frac{\varphi^f}{f!}} + \frac{V-1}{V} \frac{\kappa_b\bar{\varphi} + \bar{\varphi} \sum_{f=3}^q \kappa_b^{f/2} \kappa_f \frac{\varphi^{f-2}}{(f-2)!}}{1 + \sqrt{\kappa_b}\bar{\varphi} + \kappa_b \bar{\varphi}\varphi + \bar{\varphi} \sum_{f=3}^q \kappa_b^{f/2} \kappa_f \frac{\varphi^{f-1}}{(f-1)!}} \, , \label{eq:SaddlePointEq-1}\\
\frac{\varphi}{q}
& = & \frac{V-1}V \frac{\sqrt{\kappa_b} + \kappa_b \varphi + \sum_{f=3}^q \kappa_b^{f/2} \kappa_f \frac{\varphi^{f-1}}{(f-1)!}}{1 + \sqrt{\kappa_b} \bar{\varphi} + \kappa_b \bar{\varphi}\varphi + \bar{\varphi}\sum_{f=3}^q \kappa_b^{f/2} \kappa_f \frac{\varphi^{f-1}}{(f-1)!}} \, .
\end{eqnarray}
Note that with this approximation, every dependence upon ${\mathbf x}_0$ disappears.
In order to simplify the calculations and proceed analytically, we now take the thermodynamic limit $V \gg 1$ and neglect terms ${\mathcal O}(1/V)$.
This leads to
\begin{eqnarray}
\Xi 
& \simeq &
\exp\bigg[ -V \frac{\varphi\bar{\varphi}}q + V\log\bigg( 1 + \sqrt{\kappa_b}\bar{\varphi} + \kappa_b\bar{\varphi}\varphi + \bar{\varphi}\sum_{f=3}^q \kappa_b^{f/2} \kappa_f \frac{\varphi^{f-1}}{(f-1)!} \bigg) \bigg] \, , \label{eq:MFPartitionFunction}\\
\frac{\bar{\varphi}}q
& = &
\frac{\kappa_b \bar{\varphi} + \bar{\varphi} \sum_{f=3}^q \kappa_b^{f/2} \kappa_f \frac{\varphi^{f-2}}{(f-2)!}}{1 + \sqrt{\kappa_b} \bar{\varphi} + \kappa_b \bar{\varphi}\varphi + \bar{\varphi}\sum_{f=3}^q \kappa_b^{f/2} \kappa_f \frac{\varphi^{f-1}}{(f-1)!}} \, , \label{eq:SP1}\\
\frac{\varphi}q
& = &
\frac{\sqrt{\kappa_b} + \kappa_b\varphi + \sum_{f=3}^q \kappa_b^{f/2} \kappa_f \frac{\varphi^{f-1}}{(f-1)!}}{1 + \sqrt{\kappa_b}\bar{\varphi} + \kappa_b\bar{\varphi}\varphi + \bar{\varphi}\sum_{f=3}^q \kappa_b^{f/2} \kappa_f \frac{\varphi^{f-1}}{(f-1)!}} \, . \label{eq:SP2}
\end{eqnarray}
At this point, one can solve (eventually through numerical methods) Eqs.~\eqref{eq:SP1} and~\eqref{eq:SP2} and plug the solutions for $\varphi$ and $\bar{\varphi}$ back into Eq.~\eqref{eq:MFPartitionFunction} to have the mean-field estimate of $\Xi$. 
The saddle-point equations can also be recast in an equivalent form that will be useful in the next Section.
By multiplying both sides of Eq.~\eqref{eq:SP2} by $\bar{\varphi}$ it is easy to see that
\begin{equation}\label{eq:RHS1}
\left( 1 - \frac{\varphi\bar{\varphi}}q \right)^{-1} = 1 + \sqrt{\kappa_b} \bar{\varphi} + \kappa_b \bar{\varphi}\varphi + \bar{\varphi} \sum_{f=3}^q \kappa_b^{f/2} \kappa_f \frac{\varphi^{f-1}}{(f-1)!} \, .
\end{equation}
We can now do the same with Eq.~\eqref{eq:SP1} by multiplying both sides by $\varphi$ and get:
\begin{equation}\label{eq:RHS2}
1-\frac{\varphi\bar{\varphi}}q = \frac{1 + \sqrt{\kappa_b} \,  \bar{\varphi} - \bar{\varphi} \sum_{f=3}^q \kappa^{f/2}_b\kappa_f\frac{(f-2)\varphi^{f-1}}{(f-1)!}}{1 + \sqrt{\kappa_b} \bar{\varphi} + \kappa_b \bar{\varphi}\varphi + \bar{\varphi}\sum_{f=3}^q \kappa_b^{f/2} \kappa_f \frac{\varphi^{f-1}}{(f-1)!}} \, .
\end{equation}
Then, by equating the right-hand sides of Eqs.~\eqref{eq:RHS1} and~\eqref{eq:RHS2} we get the following relation:
\begin{equation}\label{eq:UsefulIdentity}
\sqrt{\kappa_b} = \sum_{f = 3}^q \kappa_b^{f/2} \kappa_f \frac{(f-2)}{(f-1)!} \varphi^{f-1} \, .
\end{equation}
%

\subsection{Legendre transform}\label{sec:LegendreTransform}
Now that we have an estimate of $\Xi$, 
we need to switch back to the ``more convenient'' ensemble where the total node density $\rho$ and the branch-node fractions $\phi_f = N_f/N$ (with $f \geq 3$) are fixed. 
To this purpose, we employ the Legendre-transform technique, which requires to find a way to express the fugacities in terms of $\rho$ and $\{ \phi_f \}$.
First, since only a single tree is present, we have that $N = N_b + 1$.
Therefore, since we neglect terms ${\mathcal O}(1/V)$, the total node density can be approximated by the total {\it bond} density.
Thus, we can write:
\begin{equation}\label{eq:ApproximateRho-1}
\rho \simeq \frac{\kappa_b}{V} \frac{\partial \log(\Xi)}{\partial \kappa_b} 
= \frac{\sqrt{\kappa_b} \frac{\bar{\varphi}}2 + \kappa_b \varphi\bar{\varphi} + \frac{\bar{\varphi}}2 \sum_{f=3}^q \kappa_b^{f/2} \kappa_f \frac{f}{(f-1)!} \varphi^{f-1}}{1 + \sqrt{\kappa_b} \bar{\varphi} + \kappa_b \bar{\varphi}\varphi + \bar{\varphi} \sum_{f=3}^q \kappa_b^{f/2} \kappa_f \frac{\varphi^{f-1}}{(f-1)!}} \, .
\end{equation}
Using Eq.~\eqref{eq:UsefulIdentity}, it is easy to see that:
\begin{equation}\label{eq:ApproximateRho-2}
(1-\rho)^{-1} = 1 + \sqrt{\kappa_b}\bar{\varphi} + \kappa_b\bar{\varphi}\varphi + \bar{\varphi}\sum_{f=3}^q \kappa^{f/2}_b\kappa_f \frac{\varphi^{f-1}}{(f-1)!} \, ,
\end{equation}
and, because the r.h.s. terms of Eq.~\eqref{eq:RHS1} and Eq.~\eqref{eq:ApproximateRho-2} are identical, we conclude that:
\begin{equation}\label{eq:ApproximateRho-3}
\rho = \frac{\varphi\bar{\varphi}}q \, .
\end{equation}
The $f$-functional branch-node density is instead given by
\begin{equation}\label{eq:ApproximatePhiF}
\frac{N_f}V = \frac{\kappa_f}V \frac{\partial \log (\Xi)}{\partial \kappa_f} 
= \frac{\bar{\varphi} \, \kappa_b^{f/2} \, \kappa_f \, \frac{\varphi^{f-1}}{(f-1)!}}{1 + \sqrt{\kappa_b} \, \bar{\varphi} + \kappa_b \, \bar{\varphi}\varphi + \bar{\varphi}\sum_{f'=3}^q \kappa_b^{f'/2} \, \kappa_f' \, \frac{\varphi^{f'-1}}{(f'-1)!}} \, .
\end{equation}
At this point, we can obtain an expression for $N_2/V$, namely
\begin{equation}\label{eq:BifunctionalPoints}
\frac{N_2}V =
\rho - \sum_{f=3}^{q}(f-1)\frac{N_f}{V} = 
\frac{\sqrt{\kappa_b} \frac{\bar{\varphi}}2 + \kappa_b \varphi\bar{\varphi} -\frac{\bar{\varphi}}2 \sum_{f=3}^q \kappa_b^{f/2} \kappa_f \frac{(f-2)}{(f-1)!}\varphi^{f-1}}{1 + \sqrt{\kappa_b}\bar{\varphi} + \kappa_b \bar{\varphi}\varphi + \bar{\varphi} \sum_{f=3}^q \kappa_b^{f/2} \kappa_f \frac{\varphi^{f-1}}{(f-1)!}} \, ,
\end{equation}
where, we recall, the first equality is true since we keep neglecting terms ${\mathcal O}(1/V)$.
Then, again by using Eq.~\eqref{eq:UsefulIdentity}, we get:
\begin{equation}
\frac{N_2}V = \frac{\kappa_b\varphi\bar{\varphi}}{1 + \sqrt{\kappa_b}\bar{\varphi} + \kappa_b \bar{\varphi}\varphi + \bar{\varphi}\sum_{f=3}^q \kappa_b^{f/2} \kappa_f \frac{\varphi^{f-1}}{(f-1)!}} = \kappa_b \, q \, \rho (1-\rho) \, .
\end{equation}
Therefore, we are able to express the bond fugacity as a function of $\phi_2$ and $\rho$, {\it i.e.}
\begin{equation}\label{eq:BondFugacity}
\kappa_b = \frac{\phi_2}{q(1-\rho)} \, .
\end{equation}
We can now repeat a similar procedure for all the $\kappa_f$'s.
To this end, it is useful to notice that
\begin{equation}
\frac{N_1}{V} = \sum_{f=3}^q (f-2) \frac{N_f}V = (1-\rho)\sqrt{\kappa_b} \, \bar{\varphi} \implies \bar{\varphi} = \frac{\rho\phi_1}{(1-\rho)\sqrt{\kappa_b}} \, .
\end{equation}
Then, by simple manipulations, we obtain:
\begin{equation}\label{eq:BranchingPointFugacity}
\kappa_f = (f-1)! \, \phi_f \frac{\phi^{f-2}_1}{\phi^{f-1}_2} \, .
\end{equation}
We now have all the elements to compute the Legendre transform analytically, and the entropy per lattice site
\begin{equation}\label{eq:SperLatticeSite}
\frac{s_{\rho}^{{\mathbf x}_0}(\{\phi_f\})}{k_B} = \frac1V \log(\Xi) 
- \rho\log(\kappa_b) - \rho\sum_{f=3}^q \phi_f \log(\kappa_f) = -\rho - \log(1-\rho)  - \rho\log(\kappa_b) - \rho\sum_{f=3}^q \phi_f \log(\kappa_f) \, .
\end{equation}
With Eqs.~\eqref{eq:BondFugacity} and~\eqref{eq:BranchingPointFugacity}, Eq.~\eqref{eq:SperLatticeSite} becomes equivalent to Eq.~\eqref{eq:MFFreeEnergy} in main text.
This concludes the derivation.

\section{A note on [Wagner, Erdemci-Tandogan, Zandi, J. Phys.: Condensed Matter (2015), Ref.~\cite{Wagner2015}]}\label{sec:CommentOnZandiWork}
In Ref.~\cite{Wagner2015}, Wagner {\it et al.} proposed a field-theoretic approach based on the $O(n\to0)$-model to describe polydisperse solutions of branching polymers {\it without} loops. 
However, as we are going to show here, an overlooked technical detail in their derivation implies that the statistical weight of looped conformations in the field-theoretic expression for the partition function of the system is strictly $\neq 0$, namely loops are effectively present.

Essentially, the work of Wagner {\it et al.}\footnote{Notice that similar ideas can be found in works~\cite{LubenskyIsaacson1979,Safran2002}.} is based on the idea of introducing apt external fields in the Hamiltonian of the magnet such that the ``$n \to 0$''-limit becomes equivalent to considering only conformations of self-avoiding branched polymers.
Rephrasing in the same language and notation of Sec.~\ref{sec:FieldTheoryCompleteDerivation}, we denote here with $J_1({\mathbf x})$ and $J_3({\mathbf x})$ the value of the external fields at lattice site of spatial coordinate $\mathbf x$.
Then, the formalism by Wagner {\it et al.} is equivalent to introducing the following trace properties (to be compared to our rules (I)-(VI) presented in Sec.~\ref{sec:FieldTheoryCompleteDerivation}) for the spins:
\begin{itemize}
\item[(I)]
Spins on different sites are independent of each other.
\item[(II)]
$\langle 1 \rangle = 1$.
\item[(III)]
$\langle S_i({\mathbf x}) \rangle = 0$.
\item[(IV)]
$\langle S_i({\mathbf x}) S_j({\mathbf x}) \rangle = \delta_{ij}$.
\item[(V)]
$\langle S_{i_1}({\mathbf x}) S_{i_2}({\mathbf x}) \dots S_{i_p}({\mathbf x}) \rangle = 0$ for $p \geq3$.
\item[(VI)]
$\langle S_{i}({\mathbf x}) J_1({\mathbf x}) \rangle = \delta_{i,1}$.
\item[(VII)]
$\langle S_{i_1}({\mathbf x}) S_{i_2}({\mathbf x}) S_{i_3}({\mathbf x}) J_3({\mathbf x}) \rangle = \delta_{i_1, i_2, i_3, 1}$.
\end{itemize}
At the same time, all the other traces involving the external fields $J_1$ and $J_3$ are set $=0$.
In terms of these rules, the Wagner-Erdemci-Tandogan-Zandi grand canonical partition function $\Xi_{\rm WETZ}$ of the system is given by
\begin{equation}\label{eq:WETZ-GrandCanonical}
\Xi_{\rm WETZ} = \lim_{n \to 0} \bigg\langle e^{\frac12 \sum_{{\mathbf x}, {\mathbf x}'} {\mathbf S}({\mathbf x}) \cdot {\mathbf S}({\mathbf x}') \Delta({\mathbf x}, {\mathbf x}')}\prod_{\mathbf x} \bigg( 1 + J_1({\mathbf x}) + J_3({\mathbf x}) \bigg) \bigg\rangle \, ,
\end{equation}
where all fugacities have been set to $=1$ for simplicity.
Now notice that, due to the trace rules, the exponential term in Eq.~\eqref{eq:WETZ-GrandCanonical} can be Taylor-expanded and only terms up to the first order need to be retained, since the rest of higher order terms contribute ${\mathcal O}(n)$ and, therefore, disappear in the $n \to 0$ limit.

\begin{figure}
\includegraphics[scale=0.40]{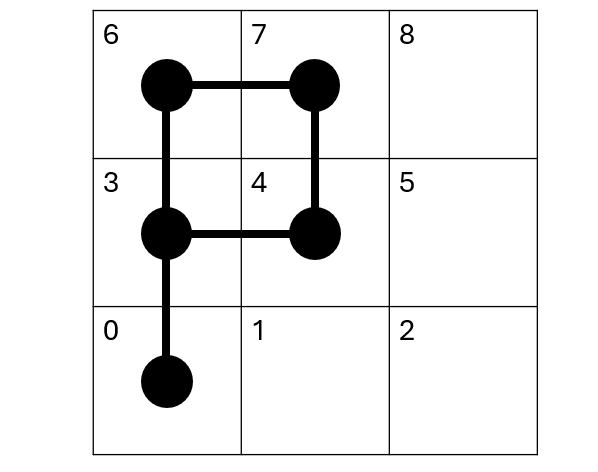}
\caption{
Illustrative conformation of a branched polymer containing a loop on a $3\times3$ square lattice.
Sites of the lattice are numbered from $0$ to $8$.
}
\label{fig:LoopExample}
\end{figure}

Similarly to Sec.~\ref{sec:On->0model-ii}, every term in the sum can be illustrated with a graphical example.
Thus a term $\sim J_1({\mathbf x})$ corresponds to the presence of an end-point on lattice site ${\mathbf x}$, while a term $\sim J_3({\mathbf x})$ indicates the presence of a branch-node (with functionality $=3$).
Then, analogously to our model (see Sec.~\ref{sec:On->0model-i}), a term ${\mathbf S}({\mathbf x}) \cdot {\mathbf S}({\mathbf x}')$ represents an {\it undirected} bond between lattice positions ${\mathbf x}$ and ${\mathbf x}'$. 
Consider now the conformation depicted in Fig.~\ref{fig:LoopExample}, whose weight corresponds to the expression:
\begin{equation}\label{eq:LoopedExample}
\sum_{\substack{j_1, j_2, j_3 \\ j_4, j_5}} \langle J_1({\mathbf x}_0) \, S_{j_1}({\mathbf x}_0) S_{j_1}({\mathbf x}_3) \, J_3({\mathbf x}_3) \, S_{j_2}({\mathbf x}_3) S_{j_2}({\mathbf x}_4) \, S_{j_3}({\mathbf x}_4) S_{j_3}({\mathbf x}_7) \, S_{j_4}({\mathbf x}_7) S_{j_4}({\mathbf x}_6) \, S_{j_5}({\mathbf x}_6) S_{j_5}({\mathbf x}_3) \rangle \, .
\end{equation}
By factorizing terms referring to different sites and after using the trace properties, it is easy to see that the weight of the conformation is $\neq 0$ despite the conformation contains a loop.
In other words, the choice of Wagner {\it et al.} of breaking the $O(n)$-symmetry by choosing a privileged component for the external fields (the component $1$, see Eqs.~(A.20) and~(A.21) in the appendix of Ref.~\cite{Wagner2015}) is not harmless, and ultimately leads to polymer configurations containing loops:
for further validation, notice that the partition function for branching polymers of Wagner {\it et al.} (see Eq.~(A.22) in their work~\cite{Wagner2015}) is in essence identical to the partition function of the model by Lubensky and Isaacson~\cite{LubenskyIsaacson1979} (see Eqs.~(A11),~(A12) and~(A13) therein) who, as pointed out by the same authors, includes indeed loops.

\section{Enumeration of spanning trees}\label{sec:EnumerationSpanningTrees}
With $\rho = 1$, Eq.~(5) 
in main text gives the entropy of spanning trees of $V$ nodes and given number of branch-nodes $N_1, N_2, ..., N_q$.
Thus, by multiplying by $V$ both sides of Eq.~(5) 
in main text and then exponentiating, we get the mean-field estimate of the total number of such trees.
In order to obtain the {\it total} number of spanning trees (\textit{i.e.} with unrestricted node functionality), this quantity must be summed over {\it all possible} values $N_1, N_2, \dots, N_q$ with constraints Eqs.~(1) and~(2) 
in main text.

As, however, this procedures is quite unpractical, we consider the following, and much simpler, alternative.
We switch back to the ensemble where the total number of nodes is fixed whereas every $N_f$ can fluctuate under control of the appropriate external fugacity $\kappa_f$.
Similarly to Sec.~\ref{sec:LegendreTransform}, we perform a Legendre transform and, by indicating with ${\tilde s} = {\tilde s}_{\rho}(\kappa_3, \dots, \kappa_q) \equiv {\tilde s}_{\rho}(\{\kappa_f\})$ the entropy per site {\it in this new ensemble}, we have the simple expression (see also Eqs.~\eqref{eq:BondFugacity} and~\eqref{eq:SperLatticeSite}):
\begin{equation}\label{eq:Road2EntropySpanningTrees}
\frac{{\tilde s}_{\rho}(\{\kappa_f\})}{k_B}
= \frac{s_{\rho}(\{\phi_f\})}{k_B} + \rho\sum_{f = 3}^q \phi_f\log(\kappa_f)
= \rho\log\left(\frac{q}e\right) - (1-\rho)\log(1-\rho) -\rho\log(\phi_2) \, ,
\end{equation}
where the $\phi_f$'s are functions of the $\kappa_f$ via Eq.~\eqref{eq:BranchingPointFugacity}.
Now, the entropy per lattice site of unweighted spanning trees in the thermodynamic limit is simply given by
\begin{equation}\label{eq:EntropySpanningTrees}
\frac{{\tilde s}_{\rho=1}(\kappa_3=1, \dots, \kappa_q=1)}{k_B} = \log\left( \frac{q}e \right) - \log(\phi_2) \, .
\end{equation}
In Table~\ref{tab:ComparisonExactMeanField}, we compare this result with some {\it exact} values for different lattice architectures reported in Refs.~\cite{Wu1977, Shrock2000}.
From the comparison, it is apparent how our MF approach becomes more and more accurate with increasing lattice coordination number $q$.

\begin{table}[h]
\centering
\begin{tabular}{cccc}
\hline
$\,\,\,$ $q$ $\,\,\,$ & $\,\,\,$ Exact~\cite{Wu1977, Shrock2000} $\,\,\,$ & $\,\,\,$ Bethe approximation~\cite{deLosRios2000} / Deviation $\,\,\,$ & $\,\,\,$ Mean-field (this work) / Deviation $\,\,\,$ \\
\hline
3 & 0.8076648 & 0.836988 / 3.6\% & 0.979986 / 21.3\% \\
4 & 1.1662436 & 1.2164 / 4.3\% & 1.36474 / 17.0\% \\
6 (Triangular) & 1.6153297 & 1.69108 / 4.7\% & 1.79116 / 10.9\% \\
6 (Cubic) & 1.6741481 & 1.69108 / 1.0\% & 1.79116 / 7.0\% \\
8 & 2.0000 & 2.00777 / 0.4\% & 2.07943 / 4.0\% \\
10 & 2.243 & 2.24691 / 0.2\% & 2.30258 / 2.7\% \\
\hline
\end{tabular}
\caption{
Entropy of spanning trees with unrestricted node functionality per lattice site and for regular lattices of coordination number $q$.
Column 2 reports the computed exact values from Refs.~\cite{Wu1977,Shrock2000}.
Column 3 is the result of the Bethe approximation for spanning trees by De Los Rios et al.~\cite{deLosRios2000} (see Eq.~(5) therein) and its relative deviation from the exact value.
Column 4 is the result of the present MF prediction (Eq.~\eqref{eq:EntropySpanningTrees}) and its relative deviation from the exact value.
}
\label{tab:ComparisonExactMeanField}
\end{table}
%

\section{Monte Carlo algorithm for spanning trees}\label{sec:MCalgos}
As explained in main text, the Amoeba algorithm~\cite{SeitzKlein1981,Rosa2016a,Rosa2016b,VanDerHoek2024} is not suitable for simulating lattice trees when the volume fraction is $=1$.
Therefore, in the particular case of spanning trees, we have devised a new Monte Carlo algorithm to sample trees based on the canonical distribution Eq.~(10) 
in main text, therefore in the ensemble where nodes' functionality is restricted to $\leq 3$ and such that traditional enumerating algorithms like~\cite{Broder1989,Aldous1990} are not applicable.

Starting from an arbitrary spanning tree conformation, the algorithm works as follows:
\begin{enumerate}
\item
Choose randomly a ``$f=1$''-node.
\item
Delete the bond between the node and the rest of the tree.
\item
Reconnect the node to one of its nearest neighbors picked at random.
\end{enumerate}
If the chosen nearest-neighbor node is already a branch-node, then the move is automatically rejected.
Otherwise, in order to ensure detailed balance, the move to change state is accepted with the following probability according to the Metropolis-Hasting rule~\cite{metropolis}:
\begin{equation}\label{eq:Acceptance}
\alpha_{i\rightarrow f} = \min\bigg\{ 1, \frac{N^{(i)}_1}{N^{(f)}_1}\exp\bigg[ {\beta\mu_3(N^{(f)}_3 - N^{(i)}_3}) \bigg] \bigg\} \, ,
\end{equation}
where the superscripts $(i)$ and $(f)$ refer to quantities measured in the {\it initial} conformation (before the move) and in the {\it final} one (after the move).
It is very easy to check that this algorithm can never lead to the formation of closed loops, and it also guarantees that the maximum functionality of a node in any conformation is always $\leq 3$.
The algorithm can be easily extended to systems with higher maximum functionality: it suffices to properly modify Eq.~\eqref{eq:Acceptance} by introducing additional chemical potentials.

\section{Generalization to poor-solvent conditions}\label{sec:PoorSolventConditions}
The SAT model considered in this work applies to polymers in so called good-solvent conditions~\cite{RubinsteinColby}, namely to the situation where the effective interaction between the polymer and the surrounding solvent dominates over monomer-monomer ({\it i.e.} node-node) interactions.
Following a procedure similar to the one presented in Sec.~\ref{sec:FieldTheoryCompleteDerivation}, it is however possible to obtain the mean-field expression for the free energy of the system also in the opposite situation of poor-solvent conditions~\cite{RubinsteinColby}, namely when an effective {\it attractive} interaction between nodes occupying nearest-neighbor lattice sites is considered in the system.

Indicating with $N_{mm}$ the total number of these node-node neighboring pairs and with $\epsilon > 0$ the intensity of the attractive interaction, the energy associated with a given configuration $\mathcal C$ is written as $-\epsilon N_{mm}({\mathcal C})$.
With reference to Eq.~\eqref{eq:GrandCanonicalZ}, the corresponding grand canonical partition function of the system, denoted with $\Xi_{ps}$, is then given by
\begin{equation}\label{eq:GCPartFunct-PoorSolvent}
\Xi_{ps} = \sum_{\{ {\mathcal C} \}_{{\mathbf x}_0}} \bigg( \kappa^{N_b({\mathcal C})}_b \, \prod_{f=3}^q \kappa_f^{N_f({\mathcal C})} \bigg) \, e^{\beta \epsilon N_{mm}(\mathcal{C})} \, .
\end{equation}
By generalizing Eq.~\eqref{eq:ZDiagramaticExpansion}, one can show that
\begin{equation}\label{eq:ZPoorSolvent}
\Xi_{ps} = \lim_{n \to 0} \int {\mathcal D} \psi \, e^{-\frac12 \sum_{ {\mathbf x}, {\mathbf x'}} \psi({\mathbf x}) \Delta^{-1}({\mathbf x}, {\mathbf x'}) \psi({\mathbf x'})} \, \, \, \bigg\langle S_1({\mathbf x}_0) e^{\sqrt{\beta \epsilon} \, \psi({\mathbf x}_0)} \prod_{{\mathbf x}, {\mathbf x'}} \bigg( 1 + \kappa_b \, e^{\beta\epsilon} \, \bar{{\mathbf S}}({\mathbf x}) \cdot {\mathbf S}({\mathbf x'}) \, e^{\sqrt{\beta\epsilon} \, \psi(\bold{x'})} \, \Delta({\mathbf x}, {\mathbf x'}) \bigg) \bigg\rangle
\end{equation}
where $\psi({\mathbf x})$ is a scalar field in correspondence of site $\mathbf x$ and ${\mathcal D} \psi = (2\pi)^{-V/2}(\det{\Delta})^{-1/2} \prod_{\mathbf x} \psi({\mathbf x})$ is the associated measure (notice that the introduction of a scalar field is a standard technique to count the number of nearest-neighbor interactions in lattice models, see~\cite{Orland1996,Marcato2023,Marcato2024}).
Starting from Eq.~\eqref{eq:ZPoorSolvent}, the procedure to obtain the mean-field expression for the free energy of the system is exactly the same as the one presented in Sec.~\ref{sec:FieldTheoryCompleteDerivation}.
Finally, the resulting free-energy per lattice site, which we denote with $\omega$, reads:
\begin{equation}\label{eq:FreeEnergyPoorSolvent}
\beta \omega = -\frac{q}2 \beta\epsilon \rho^2 - \rho \ln \bigg( \frac{q}e \bigg) + (1-\rho)\log(1-\rho) + \rho\sum_{f = 1}^q \phi_f \ln[\phi_f (f-1)!] + \rho \beta\epsilon \, .
\end{equation}
Interestingly, there are two immediate and non-trivial consequences from Eq.~\eqref{eq:FreeEnergyPoorSolvent}.
First, it is clear that the branching probability $\langle \phi_f \rangle$ is independent of the interaction $\epsilon$, and then equal to the result in good-solvent presented in main text.
Second, the phase diagram for the collapse transition of a self-avoiding branching polymer is predicted at the mean-field level to be no different from the one of a linear chain, a feature that was already observed in previous works~\cite{Nemirovsky1992,deLosRios2000}: this is easily seen as all linear terms (including the branching term) cancel by imposing the known condition for the collapse transition, $\partial^2\omega / \partial\rho^2 = 0$.


\clearpage

\section*{Supplementary figures}

\clearpage
\begin{figure}
\includegraphics[width=1.0\textwidth]{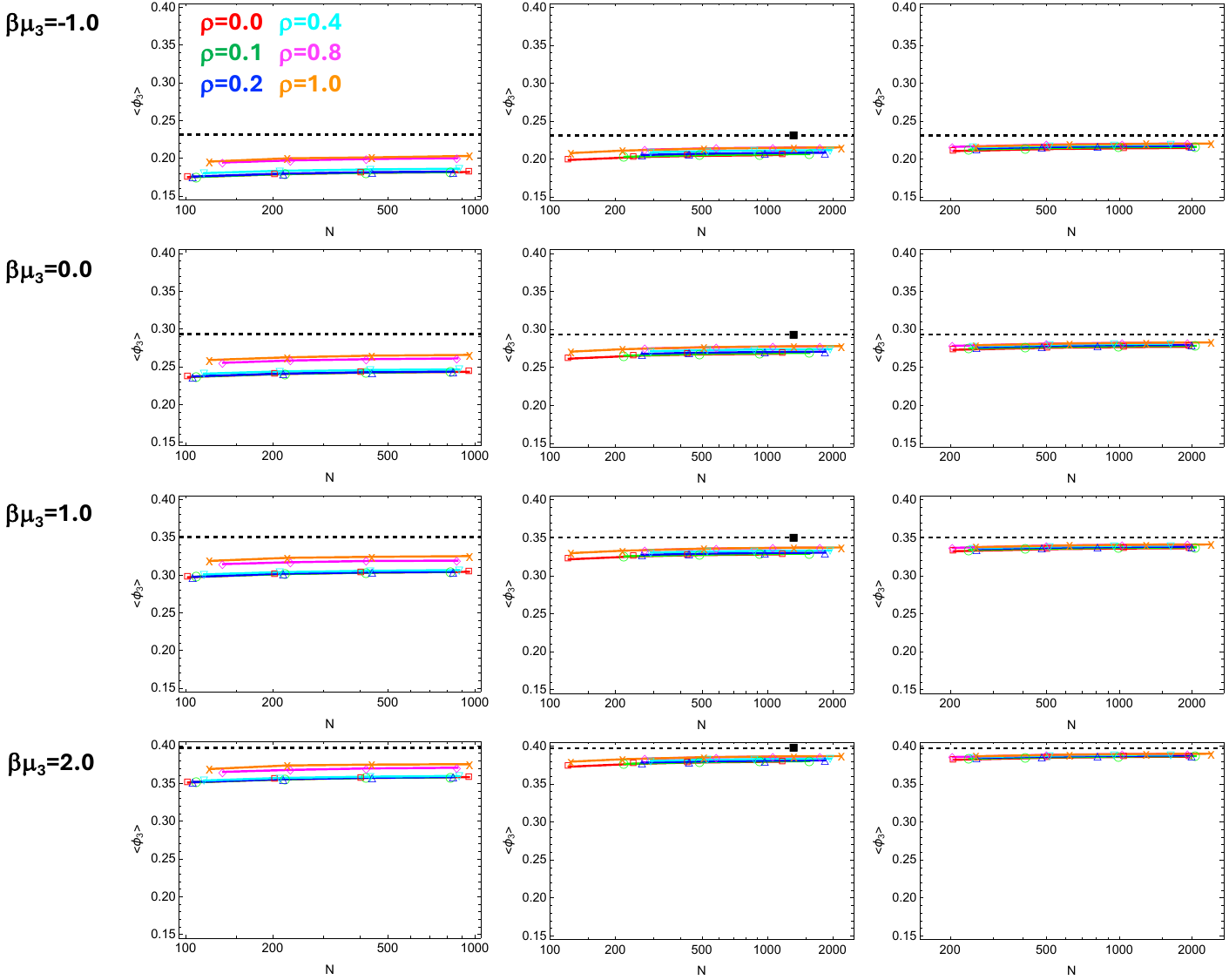}
\caption{
SAT's in $d=2,3,4$ spatial dimensions (panels from left to right).
Mean fraction of branch-nodes, $\langle \phi_3\rangle$, as a function of the total number of nodes $N$.
Dashed horizontal lines correspond to the mean-field Eq.~(9) 
in main text.
Different colors/symbols are for different node fractions $\rho=0.0,0.1,0.2,0.4,0.8,1.0$ (same color/symbol codes as in Fig.~3 
in main text), panels from top to bottom are for different values of the branch chemical potential $\beta\mu_3 =-1,0,1,2$ (see legend).
The filled square ($\blacksquare$) symbols for $d=3$ denote results for computer simulations of {\it ideal} lattice trees ({\it i.e.} no excluded volume).
}
\label{fig:2Ddata}
\end{figure}

\end{document}